\documentclass[preprint,12pt]{elsarticle}
\usepackage{tabularx}
\usepackage{lineno,hyperref}
\modulolinenumbers[5]
\usepackage{amsmath,amsfonts}
\usepackage{algorithmic}
\usepackage{algorithm}
\usepackage{array}
\usepackage{float}        
\usepackage{threeparttable}
\usepackage{xcolor}
\usepackage[caption=false,font=normalsize,labelfont=sf,textfont=sf]{subfig}
\usepackage{textcomp}
\usepackage{stfloats}
\usepackage{soul} 
\usepackage{url}
\usepackage{verbatim}
\usepackage{graphicx}
\usepackage{tabularx,booktabs,textcomp}
\usepackage{amsthm} 
\usepackage{longtable}
\usepackage{tabularray}

\usepackage{amsmath,amssymb,amsthm}

\journal{Elsevier}

\begin{document}

\begin{frontmatter}

\title{Physics-Informed Neural Network-Based Control for Grid-Forming Converter's Stability Under Overload Conditions}


\author[1,2]{Abhay Kumar}

\author[2]{Dushyant Sharma}

\author[1]{Mayukha Pal\corref{mycorrespondingauthor}}
\ead{mayukha.pal@in.abb.com}
\cortext[mycorrespondingauthor]{Corresponding author}


\affiliation[1]{organization={ABB Ability Innovation Center},
     addressline={Asea Brown Boveri Company},
    city={Hyderabad},
    postcode={500084},
    state={Telangana},
    country={India}}    

\affiliation[2]{organization={Department of Electrical Engineering},
    addressline={Indian Institute of Technology (ISM)},
    city={Dhanbad},
    postcode={826004},
    state={Jharkhand},
    country={India}}

\begin{abstract}

{Grid-forming converters (GFCs) are crucial for frequency and voltage stability in modern power systems. However, their performance under overload conditions remains a challenge. This paper highlights the limitations of existing approaches in managing DC source saturation and AC current limits, emphasizing the need for improved control strategies to ensure system stability. This paper proposes a control strategy based on a physics-informed neural network (PINN) to improve GFC performance under overloaded conditions, effectively preventing switch failures and mitigating DC source saturation. This approach outperforms conventional methods by maintaining stable voltage and frequency, even under significant load increase where traditional droop control alone proves inadequate. The post-disturbance operating point of GFCs remains unchanged using PINN-based control with an improvement of 0.245 Hz in frequency and 0.03 p.u. in active power when compared to an already existing current limitation strategy. Additionally, it reduces peak voltage deviations during transients by 24.14\%, lowers the rate of change of frequency (ROCOF) from 0.02 Hz/s to 0.005 Hz/s, and improves the rate of change of voltage (ROCOV), keeping both within acceptable limits. These improvements significantly enhance system resilience, especially in inertia-less power networks.}
\end{abstract}
\end{frontmatter}
\section{Introduction}
\label{section:Intro}
Renewable energy sources, such as photovoltaics (PV) and batteries, integrated via grid-following inverters, face challenges like low inertia and dependence on the grid, limiting their functionality in islanded systems. Grid-forming converters, functioning as voltage sources with fixed frequency, offer a solution by enabling island operation and maintaining power balance. Advanced control strategies are essential for effectively integrating into modern power systems \cite{ref1}. Comparative studies show grid-forming converters surpass grid-following counterparts in autonomy and frequency stability \cite{ref1}. Grid-forming control strategies have been extensively studied to enhance the stability and performance of inverter-based power systems. As discussed in \cite{ref2}, droop control is a widely adopted method for managing parallel-connected inverters in standalone AC supply systems. State-space modeling techniques for autonomous microgrid inverters are explored in \cite{ref3}, along with a sensitivity analysis that aids in system analysis. Synchronverters, introduced in \cite{ref4}, emulate synchronous generators and offer improvements in harmonic distortion and the fluctuating amplitude of grid-side currents, as presented in \cite{ref5}. Equivalences between virtual synchronous machines (VSM) and frequency droop control are examined in \cite{ref6}, while virtual oscillator control (VOC), introduced in \cite{ref7}, stabilizes interconnected inverters by mimicking nonlinear oscillator dynamics. A comparative study in \cite{ref8} highlights VOC's superior stability performance in the time domain over droop control for interconnected inverters. Transmission line effects analyzed in \cite{ref9} reveal destabilizing impacts on multi-inverter systems, emphasizing the importance of properly tuning inverter control gains. Synchronous machine-inspired matching control is proposed in \cite{ref10}, while \cite{ref11} explores various GFC strategies, including droop control, VSM, matching control, and dispatchable VOC (dVOC). Among these, matching control demonstrates improved frequency stability but shows a higher rate of change of frequency (ROCOF) under sudden large load conditions. 
All the methods discussed above provide stability to the system at nominal and rated power levels of the GFC. However, the stability of GFC under conditions involving DC source saturation and AC current limits has not been extensively explored, particularly in scenarios with sudden large power demands that exceed the GFC's rated power. To address this issue, \cite{ref11} introduces a current limitation strategy that successfully stabilizes droop-controlled GFCs; however, this method affects the post-disturbance operating point of GFCs due to its threshold value being set below the rated value. Moreover, all these model-based controllers rely heavily on precise tuning and are optimized for predefined environments, limiting their adaptability to dynamic grid conditions.
\par The resilience of GFCs in inertia-less systems, particularly with appropriate DC-side and AC-side voltage and current regulation under diverse operating conditions, remains a key focus of this paper. Fully renewable power systems, increasingly reliant on inverter-based resources, face significant stability challenges, as highlighted in \cite{ref12}. Enhancements to droop control, such as integral sliding mode control, address issues like DC-side voltage collapse in GFCs, improving resilience under varying conditions \cite{ref13}. Decoupling active and reactive power in GFCs using advanced techniques like multivariable control offers superior performance, as discussed in \cite{ref14}. Model predictive control (MPC) integrated with VSM enhances system performance, reliability, and dynamic response, particularly in fluctuating grid environments \cite{ref15}. Additionally, adaptive inertia-based virtual synchronous generator designs for distributed generation (DG) inverters provide improved flexibility and performance in dynamic operating conditions \cite{ref16}. PI controllers are commonly employed in GFC control loops to manage voltage and frequency variations \cite{ref17}. However, precise tuning of PI controllers is challenging due to the complex mathematical modeling involved. To address this, data-driven control strategies have been introduced, simplifying the system modeling required for controller tuning. Reference \cite{ref18} reveals limitations in system adaptability in inertia-less microgrids. To enhance voltage and frequency stability in low-inertia microgrids, synchronous generator-integrated control loop signals have been explored \cite{ref19}. Secondary and complementary control loops using fuzzy-based controllers have also been applied to manage voltage and frequency in distributed converters \cite{ref20}\cite{ref21}. However, traditional controllers may struggle to maintain stability in weak microgrids under severe disturbances and uncertainties in distributed resources \cite{ref22}. To address these challenges, advanced intelligent control techniques have been proposed. Robustness-informed deep learning neural networks \cite{ref22} and convolutional neural networks (CNNs) \cite{ref23} have shown promising results in improving microgrid voltage and frequency stability.
\begin{table}[H] 
\setlength{\tabcolsep}{1.7pt}
\renewcommand{\arraystretch}{0.3}
\centering
\caption{\textsc{Modern Control Techniques Using Power Converters: A Comparison Between 2020 and 2024}}
\begin{threeparttable}
\begin{tabular}{|m{65pt}|m{40pt}|m{50pt}|m{35pt}|m{67pt}|m{25pt}|m{30pt}|m{30pt}|}
\hline
Controllable Device & Optimal Parameter Design & Control Features & Works in inertia-less MGs & Transition Modes & DC Saturation & AC Current Limitation & Same post disturbance operating point\\ 
\hline
Synchronous machine and multiple GFCs \cite{ref11} & [\checkmark] & $P/f$ droop & [\text{\sffamily X}] & N/A & [\checkmark] & [\checkmark] & [\text{\sffamily X}]  \\ \hline
 Multiple voltage source converters \cite{ref20} & [\checkmark] & Adaptive $P/f$, $Q/V$  & [\checkmark] & reconnect to MG & N/A & N/A & N/A \\ \hline
 Distributed converter-interfaced resources \cite{ref22} & [\checkmark] & Intelligent $P/f$, $Q/V$  & [\checkmark] & N/A & N/A & N/A & N/A \\ \hline
 DSPEVs \cite{ref25} & [\checkmark] & Adaptive intelligent $P/f$, $P/V$, $Q/V$ & [\checkmark] & reconnect to/disconnect from MG & N/A & N/A & N/A \\ \hline
 Renewable-based generations (This work) & [\checkmark] & PINN with \texorpdfstring{$P/f$}{P/f} droop & [\checkmark] & reconnect to/disconnect from MG & [\checkmark] & [\checkmark] & [\checkmark] \\ \hline
\end{tabular}
\begin{tablenotes}
\footnotesize
\item *Note: $P/f$ – Active power - frequency control; $P/V$ – Active power - voltage control; $Q/V$ – Reactive power - voltage control. N/A: Not considered or reported, [\checkmark]: Ensured, [\text{\sffamily X}]: Not ensured.
\end{tablenotes}
\end{threeparttable}
\label{Tabel 1}
\end{table}
Reference \cite{ref24} examines the characteristics of renewable energy-powered microgrids and highlights the associated control challenges, with results indicating superior cost reduction compared to conventional methods. Additionally, a resilience-guided control framework proposed in \cite{ref25} ensures seamless grid synchronization for stable voltage and frequency during unforeseen islanding conditions. TABLE I highlights various studies on data-driven controllers using voltage source converters, identifying their features and limitations.
\par Despite the progress achieved in data-driven control strategies, the studies summarized in TABLE I have not addressed DC source saturation and AC current limitations in GFCs that are necessary to protect semiconductor switches and the DC source under overload conditions. This paper addresses this gap by investigating GFC stability under sudden overloading conditions, offering an improved control strategy that considers these critical factors. 
In conclusion, the data in Table I reveal significant gaps, including
\begin{itemize}
    \item Previous research has not investigated the impact of AC current limitations in GFC during abrupt overload on the voltage and frequency stability of inertia-less microgrids using data-driven methods. 
    \item Previous research has not investigated the impact of DC source saturation in GFC during abrupt overload on the voltage and frequency stability of inertia-less microgrids using data-driven methods. 
    \item Applications of GFC with smooth grid connection and disconnection that take AC current limitation and DC source saturation into account have not been studied in previous studies using data-driven methods. 
    \item Previous research has not investigated the impact of AC current limitations and DC source saturation in GFC during abrupt overload on the voltage and frequency stability of inertia-less microgrids without influencing the post-disturbance operating point of the GFC.
\end{itemize}
To accomplish these goals, this study emphasizes the following contributions:
\begin{itemize}
    \item A data-driven control approach based on a physics-informed neural network (PINN) is proposed for GFCs to enhance smooth transitions during islanding and grid-connected modes.
    \item This work incorporates the AC current limitation to protect the GFC during overload conditions without influencing the post-disturbance operating point of the GFCs.
    \item Furthermore, this work ensures the stability of the microgrid by preventing saturation of the DC source during overload conditions.
    \item This work ensures stability of GFCs in inertia-less scenarios.
\end{itemize}
 The remainder of this paper is organized as follows: Section 2 reviews the modeling of GFCs, including their control strategies. Section 3 presents the proposed Physics-Informed Neural Network (PINN) approach for enhancing GFC stability. The system under consideration, along with simulation-based analysis of various case studies, is discussed in Section 4. Finally, concluding remarks are provided in Section 5.

\section{MODELING OF GFC}
In this study, a test setup consisting of GFCs is employed to enable simulation-based validation and performance assessment. This section outlines the modeling methodology adopted for the GFCs considered in the analysis. It includes detailed representations of the power converter, DC-side control mechanisms, and AC-side control strategies.
\begin{figure}[htbp]
	\centering{\includegraphics[width=10.5cm,height=4cm]{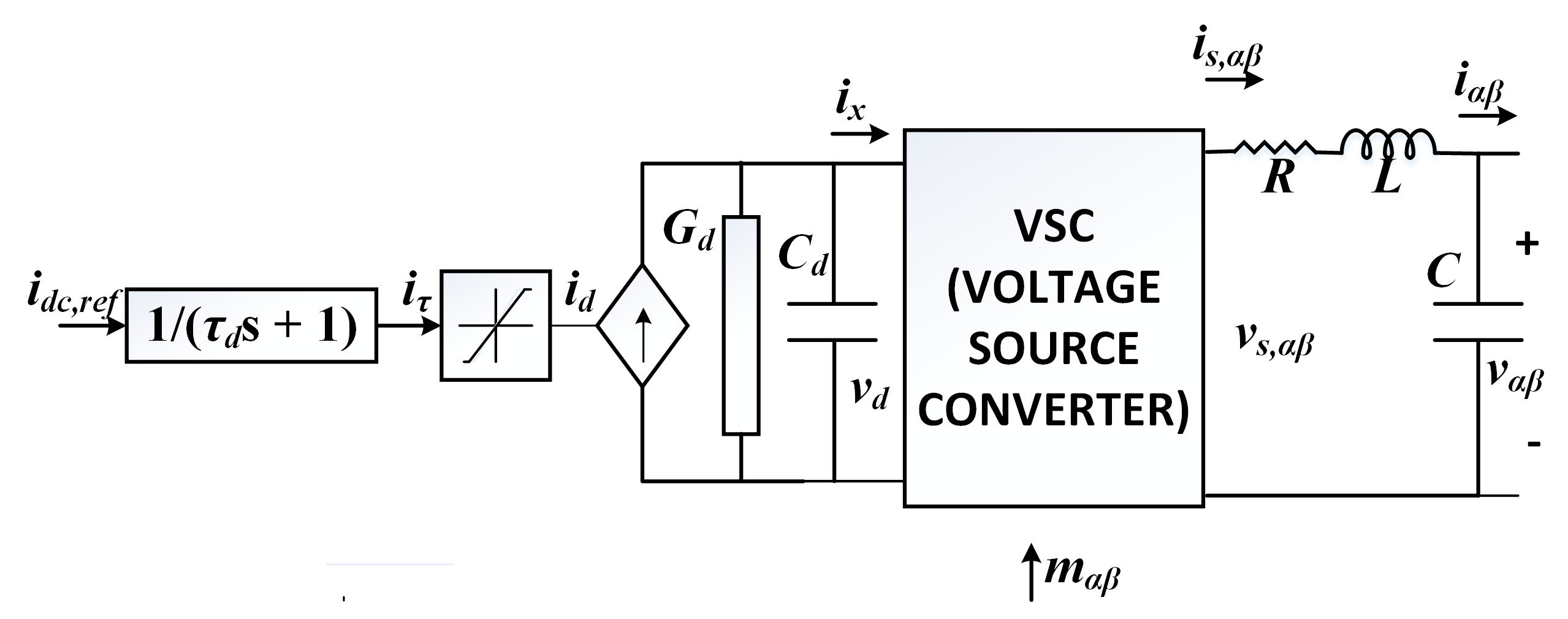}}
	\caption{Per phase equivalent model of the converter in $\alpha\beta$ coordinates }
	\label{fig.1}
\end{figure}
\subsection{Model of the converter}
First, we look at the converter model in $\alpha$$\beta$ coordinates shown in Fig.\ref{fig.1}\cite{ref11}. 
\begin{subequations}\label{eq:main}
    \begin{align}
        C_{d}\dot{v}_{d} &= i_{d} - G_{d}v_{d} - i_{x} \label{eq:1a} \\
        L\dot{i}_{s,\alpha\beta} &= v_{s,\alpha\beta} - Ri_{s,\alpha\beta} - v_{\alpha\beta} \label{eq:1b} \\
        C\dot{v}_{\alpha\beta} &= i_{s,\alpha\beta} - i_{\alpha\beta} \label{eq:1c}
    \end{align}
\end{subequations}
where $C_{d}$ is the DC link capacitance, $G_{d}$ indicates the conductance that models DC losses, $L$ refers to the filter's inductance, $C$ is the filter's capacitance, $R$ indicates the filter's resistance, $v_{d}$ refers to the DC voltage, and $i_{d}$ is the current flowing out of the controllable DC current source. $m_{\alpha\beta}$ refers to the full bridge averaged switching stage model's modulation signal. $i_{x}$ = (1/2) $m^{T}_{\alpha\beta}i_{s,\alpha\beta}$ (here $i_{s,\alpha\beta}$ indicates the ac switching node current), $v_{s,\alpha\beta}$ = (1/2) $m^{T}_{\alpha\beta}v_{d}$ (here $v_{s,\alpha\beta}$ refers to the ac switching node voltage), $i_{\alpha\beta}$ is the output current, and $v_{\alpha\beta}$ indicates the output voltage of the converter.
To accurately capture the dynamic behavior of the DC energy source, its response time is modeled using a first-order system, providing a more realistic representation for simulation and analysis purposes.

\begin{equation}
	\label{deqn_ex2}
	\tau_{d}\dot{i}_{\tau}= i_{dc,ref} - i_{\tau}
\end{equation}
where, $i_{dc,ref}$ is the DC current reference, $i_{\tau}$ refers to the DC source-provided current and $\tau_{d}$ indicates DC source time constant. Additionally, the saturation function models the DC source current limitations
\begin{equation}
	\label{deqn_ex3}
	i_{d}=\begin{cases}
		i_{\tau}, & if |i_{\tau}|< i^{d}_{max}\\
		sgn(i_{\tau})i^{d}_{max}, & if |i_{\tau}|\geq i^{d}_{max}  
		\end{cases}
\end{equation}
where $i^{d}_{max}$ is the maximum DC source current. 
In actuality, the restriction imposed by (3) is equivalent to the current constraints of a PV/wind power generating system, a DC-DC converter, or an energy storage system. Notably, to safeguard its semiconductor switches, the converter must also restrict its AC current. The study uses an aggregate model consisting of 3 converter modules (i.e. number of converter, n = 3), each rated at 500 kVA, totaling 1.5 MVA. The resultant model will have output current as 3 times $i_{\alpha\beta}$ and output voltage as $v_{\alpha\beta}$ \cite{ref11}.
\subsection{Control of GFC}
\begin{figure*}[htbp]
	\centering{\includegraphics[width=12.5cm,height=6cm]{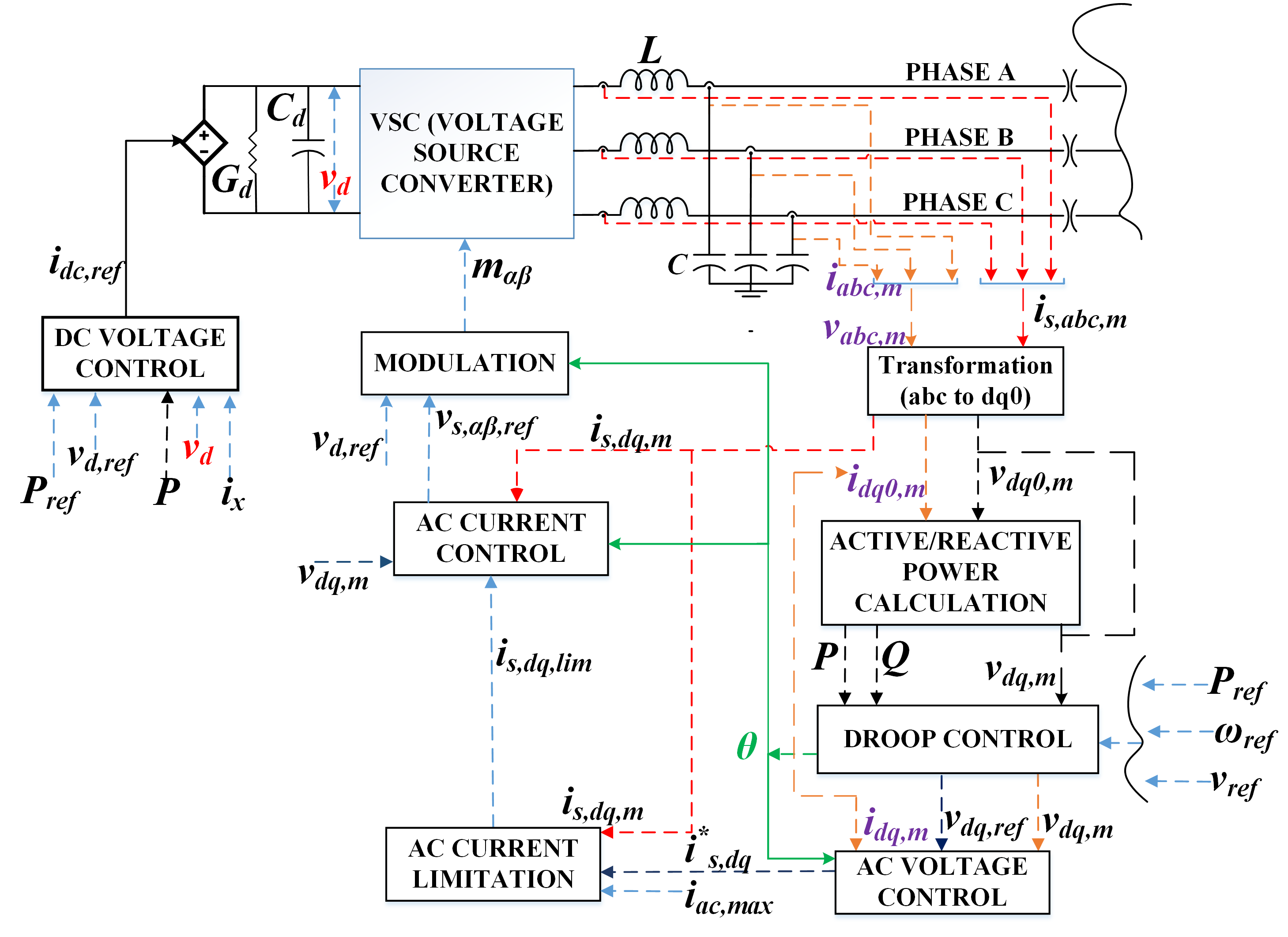}}
	\caption{Block Diagram of GFC Considered}
	\label{fig.2}
\end{figure*}
Grid-forming control strategies regulate a converter using the reference current $i_{dc,ref}$ for the DC energy source and the modulation signal $m_{\alpha\beta}$ for the DC-AC conversion stage. The block diagram of the grid forming converter considered is shown in Fig. \ref{fig.2}. The low-level cascaded control design is described as
\subsubsection{Control of DC Voltage}
The DC voltage control is managed by a current-controlled source, with a given DC current reference $i_{dc,ref}$ as
\begin{equation}
	\label{deqn_ex4}
	i_{dc,ref}= k_{d}(v_{d,ref}-v_{d})+\frac{P_{ref}}{v_{d,ref}} + (G_{d}v_{d}+\frac{v_{d}i_{x}-P}{v_{d,ref}})
\end{equation}
where, $k_{d}$ is DC voltage control gain, $v_{d,ref}$ refers the reference DC voltage, $P_{ref}$ indicates nominal active power injection, $P$ is the active power injected into the grid and $v_{d}$$i_{x}$ refers the DC power flowing into the switches. Therefore, to implement consistent DC voltage control, (4) is applied uniformly across all the control strategies discussed in this article. 
\subsubsection{Control of AC Voltage}
As illustrated in Fig. 2, our converter control system employs a modulation signal $m_{\alpha\beta}$ (in some controls like matching control operating in $\alpha\beta$ co-ordinates is helpful \cite{ref11}), which is determined by cascaded proportional-integral (PI) controllers. These controllers are implemented in $dq$-coordinates, rotating with the reference angle $\theta$, and are designed to track the voltage reference $v_{dq,ref}$.
The reference $i^{*}_{s,dq}$ for the switching node current $i_{s,dq,m}$ is calculated based on the voltage tracking error $v_{dq,ref}$-$v_{dq,m}$.\\
\begin{equation}
	\label{deqn_ex5}
	\dot{y}_{v,dq,m}= v_{dq,ref}-v_{dq,m}
\end{equation}
\begin{equation}
	\label{deqn_ex6}
	i^{*}_{s,dq}=i_{dq,m} + C{\omega}T_{2}v_{dq,m} + K_{v,p}(v_{dq,ref} - v_{dq,m}) + K_{v,i}y_{v,dq,m}
\end{equation}
where $y_{v,dq,m}$ is the integrator state, $v_{dq,ref}$ refers to the reference voltage in the $dq$ reference frame. $v_{dq,m}$ indicates the output voltage in the $dq$ reference frame. $i_{dq,m}$ is the output current in the $dq$ reference frame, C is the filter's capacitance, $\omega$ refers to the measured frequency, 
$T_{2} = \begin{bmatrix} 
0 & -1 \\ 
1 & 0 
\end{bmatrix}$
(90$^{\circ}$ rotation matrix), and $K_{v,p}$ indicates the diagonal matrix of proportional gain, $K_{v,i}$ refers to the diagonal matrix for integral gain.

\subsubsection{Limitation on AC Current}
As shown in Fig. \ref{fig.2}, AC current limitation is essential to properly protect the switches of the converter. Therefore, the reference current must remain below a specified current limit $i_{ac,max}$
\begin{equation}
	\label{deqn_ex7}
	i_{s,dq,lim}=\begin{cases}
		i^{*}_{s,dq}, & if||i_{s,dq,m}||\leq i_{ac,max}\\
	w_{i}i^{*}_{s,dq}, & if||i_{s,dq,m}||>i_{ac,max}  
		\end{cases}
\end{equation}
where $w_{i}=(i_{ac,max}/||i^{*}_{s,dq}||)$ and $i_{s,dq,lim}$ is the reference current with limitations. Limiting current is crucial for the stability margins and dynamics of GFC. In \cite{ref26a} and \cite{ref27}, the significance of current limitation is emphasized, with the maximum allowable AC current defined as 1.2 times the peak current ($I_{peak}$). This approach is implemented in the current study. 

\subsubsection{Control of AC Current}
As shown in Fig. \ref{fig.2}, the signal $i_{s,dq,lim}$ is tracked using a PI controller for the current $i_{s,dq,m}$
\begin{equation}
	\label{deqn_ex8}
 \dot{y}_{i,dq,m}=i_{s,dq,lim}-i_{s,dq,m}
\end{equation}
\begin{equation}
	\label{deqn_ex9}
	v^{*}_{dq}=v_{dq,m}+Zi_{s,dq,m}\\+K_{i,p}(i^{*}_{s,dq}-i_{s,dq,m})\\+K_{i,i}y_{i,dq,m}
\end{equation}
where $y_{i,dq,m}$ is the integrator state, $v^{*}_{dq}$ refers to the reference voltage for switching, $K_{i,p}$ denotes the diagonal matrix of proportional gain, $K_{i,i}$ denotes the diagonal matrix of integral gain, and $Z = L\omega T_{2}+RI_{2}$ (here, $I_{2}$ is the 2-D Identity Matrix, $L$ is the filter's inductance, and $R$ indicates the filter's resistance). The modulation signal, $m_{\alpha\beta}$ is
\begin{equation}
	\label{deqn_ex10}
m_{\alpha\beta}=\frac{2v_{s,\alpha\beta,ref}}{v_{d,ref}}
\end{equation}\\
where $v_{s,\alpha\beta,ref}$ is the $\alpha\beta$ coordinate image of reference voltage $v^*_{dq}$ for switching.\\ 

\subsubsection{Droop Control}
\begin{figure}[htbp]
	\centering{\includegraphics[width=10.5cm,height=3.0cm]{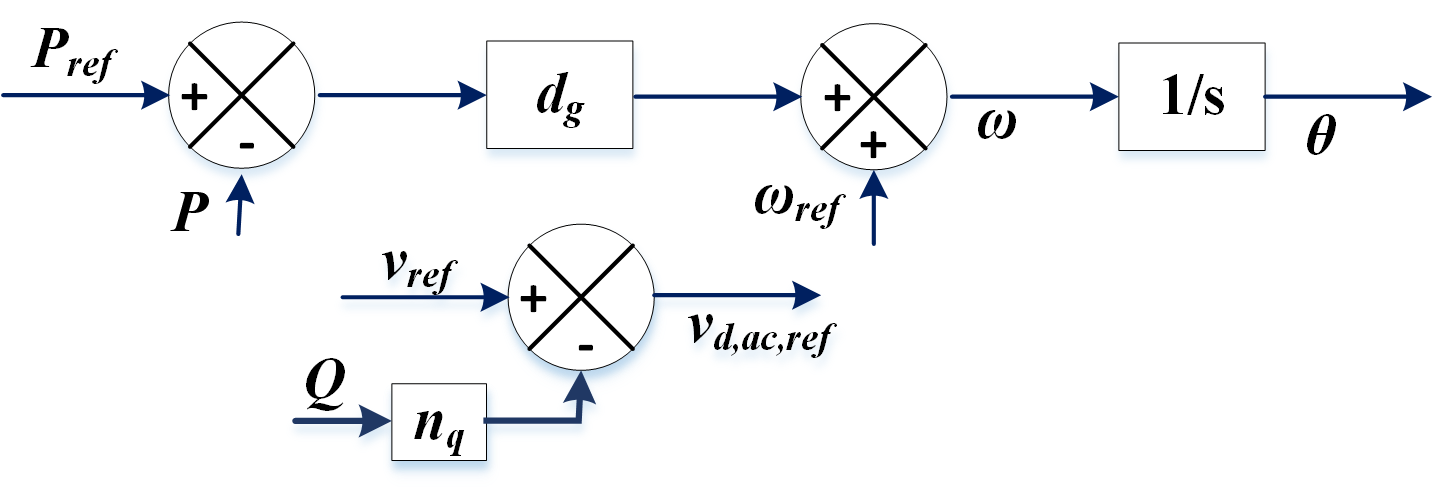}}
	\caption{Droop Control Block Diagram}
	\label{fig.3}
\end{figure}

Droop control mimics the speed droop characteristic of a synchronous machine governor, balancing power injection deviations with frequency variations
\begin{equation}
	\label{deqn_ex11}
	\dot{\theta}=\omega
\end{equation}
\begin{equation}
	\label{deqn_ex12}
	\omega=\omega_{ref}+d_{g}(P_{ref}-P)
\end{equation}
where $d_{g}$ is the active power droop coefficient and $\omega_{ref}$ indicates the reference frequency. We use the reactive power droop coefficient $n_{q}$ to control voltage as 
\begin{equation}
	\label{deqn_ex13}
	v_{d,ac,ref}=v_{ref} - n_{q}Q
\end{equation}
where $v_{d,ac,ref}$ is the direct axis reference (d-axis component of $v_{dq,ref}$ ), $v_{ref}$ refers to the reference voltage magnitude, and $Q$ indicates the reactive power. The q-axis voltage reference, $v_{q,ac,ref}$ (q-axis component of $v_{dq,ref}$ ), is set to zero. The control block diagram using droop is shown in Fig. \ref{fig.3}. 
\begin{table}[h]
    \centering
    \renewcommand{\arraystretch}{0.8}
    \setlength{\tabcolsep}{0.8pt}
    \caption{Case Study Model and Control Parameters}
    \label{tab:Table2}
    \small
    \begin{tabular}{|c|c|c|c|}
        \hline
        \multicolumn{4}{|c|}{\textbf{Single converter module}} \\
        \hline
        $S_t$ & 500 kVA & $G_{d}, C_{d}$ & 0.83 $\Omega^{-1}$, 0.008 F  \\
        $R$ & 0.001 $\Omega$ & $L$ & 200 $\mu$H  \\
        $n$ & 3 & $\tau_{dc}$ & 50 ms  \\
        $v_{d,ref}, v_{ll-rms}^{*}$ & 2.44, 1 kV &  $C$ & 300 $\mu$F\\
         $i_{max}^{d}$ & 1.2 pu(737.7A) & $i_{ac,max}$ & 1.2*$I_{peak}$(1224.7A)  \\
        \hline
        \multicolumn{4}{|c|}{\textbf{Ac current, ac voltage, and dc voltage control}} \\
        \hline
        $k_{v,p}, k_{v,i}$ & 0.52, 3.48 & $k_{i,p}, k_{i,i}$ & 1.48, 0.4  \\
        $k_{dc}$ & 1.6 $\times 10^3$ & &  \\
        \hline
        \multicolumn{4}{|c|}{\textbf{Droop control}} \\
        \hline
        $d_{g}$ & $2\pi 0.05$ rad/s & $\omega_{ref}$ & $2\pi 50$  \\ $n_{q}$ & $1.5*d_{g}$ & $P_{ref}$ & 0.3*$S_{t}$ \\
        \hline
    \end{tabular}
\end{table}
In a scenario involving grid synchronization, the reference
signals are frequency ($\omega$) and voltage ($v_{ref}$), facilitating a seamless connection of the islanding MG to the utility grid \cite{ref25}. This synchronization ensures seamless integration with the grid and facilitates stable operation under varying system conditions. The values of the parameters described so far are given in Table II \cite{ref11}.
\section{PINN FOR GFC }
\subsection{Overview of PINN}
To obtain the appropriate voltage and frequency values, the modulation index from PINN can be calculated using the $\alpha\beta$ co-ordinate image of the reference voltage signal ($v_{s,\alpha\beta,ref}$) without the need for extra controllers. Let \textit{in} $\in$ $R^{N_{in}}$ and \textit{out} $\in$  $R^{N_{out}}$ be the vectors of input and output of the PINN, which are written by

\begin{equation}
	\label{deqn_ex14}
    \begin{aligned}
\textit{in} &= \{ i_{d,m}, i_{q,m}, i_{o,m}, i_{s,d,m}, i_{s,q,m}, i_{s,o,m}, \omega, v_{d,m}, v_{q.m}, v_{o,m}, \\
&\quad \text{error} = (v_{\text{ref}} - v), v_{gd}, v_{gq}, v_{go}, P, Q, \text{ROCOF}, \text{ROCOV} \}, \\
\end{aligned}
\end{equation}

\begin{equation}
	\label{deqn_ex15}
    \begin{aligned}
\textit{out} &= \{ \hat{v}_{s,\alpha\beta,ref} \}
\end{aligned}
\end{equation}

where $i_{d,m}$, $i_{q,m}$, and $i_{o,m}$ are the $dq0$ axis components of GFC's current after filter, respectively, $i_{s,d,m}$, $i_{s,q,m}$, and $i_{s,o,m}$ are the $dq0$ axis components of GFC's current before filter, respectively. $v_{d,m}$, $v_{q,m}$, and $v_{o,m}$ are the $dq0$ axis components of GFC's voltage considering the filter's capacitor, respectively; error is the difference between the GFC's instantaneous voltage ($v$) and the reference voltage magnitude. $v_{gd}$, $v_{gq}$, and $v_{go}$ are the $dq0$ axis components of bus (to which the GFC is connected) voltage, respectively. $P$ is the output active power of the GFC, $Q$ refers to the output reactive power of the GFC, ROCOF indicates the rate of change of frequency of the GFC and ROCOV is the rate of change of voltage of the GFC.

\subsection{Key components of the PINN}
\subsubsection{Data Processing}
For model training, the \textit{in} and \textit{out} data are loaded and transformed into Pytorch tensors.
\subsubsection{Neural Network Architecture}
Each of its three hidden layers has 128 neurons. In every hidden layer, it employs the sigmoid activation function. Except for the final layer, which stays linear, it is a forward pass, meaning that every hidden layer performs a linear transformation followed by a sigmoid activation. It consists of fully connected layers, applying transformations at each layer. For a given input $x$, neural network computes:
\begin{subequations}\label{eq:main}
    \begin{align}
        h^{(0)} &= x \label{eq:16a} \\
         h^{(l+1)} &= \sigma(w^{(l)}h^{(l)} + b^{(l)})  \quad \forall{l\in[0, L-1]}\label{eq:16b} \\
        \hat{y} &= w^{L}h^{L} + b^{L}\label{eq:16c}
    \end{align}
\end{subequations}

where $L$ is the total number of hidden layers, $w^{(l)}$ and $b^{(l)}$ are the weight matrix and bias vector at layer l, $\sigma(x) = 1/(1+e^{-x})$ is the sigmoid activation function and $\hat{y}$ are the predicted reference voltages.
\subsubsection{Physics informed loss function}
It adds physical constraints to the standard mean squared error (MSE) loss. The model includes five components in its loss function, which are as follows:
\begin{itemize}
\item\textbf{Mean squared error (MSE) loss}:
It is the standard loss between predicted and actual predictions
\begin{equation}
	\label{deqn_ex17}
	L_{MSE} = 1/N \sum_{i=1}^{N}(\hat{y_{i}} - y_{i})^2
\end{equation}
where, N is the number of training samples, $\hat{y_{i}}$ is the predicted output and $y_{i}$ is the true reference voltage.
\item\textbf{Physics informed loss}: To enforce grid-forming stability, we add voltage derivative constraints
\begin{equation}
	\label{deqn_ex18}
	L_{Physics} = E[\frac{d}{dt} \hat{v}_{s,\alpha\beta,ref}+ v_{s,\alpha\beta,ref}]
\end{equation}
using automatic differentiation, we approximate ${\frac{d}{dt}\hat{v}_{s,\alpha\beta,ref}}$ $\approx$ ${\frac{\partial} {\partial t}v_{s,\alpha\beta,ref}}$ which is computed using Pytorch. The symbol ${E}[\cdot]$ denotes the expectation operator, which computes the statistical mean or average of the quantity inside the brackets.
\item\textbf{Current limitation loss}: For inverter's safety, the model penalizes currents exceeding a certain limit, for this paper the limit is considered as the 1.2 times the peak inverter current represented by $I_{peak}$. The current magnitude is given by
\begin{equation}
	\label{deqn_ex19}
	I_{total} = \sqrt{I_{d}^2 + I_{q}^2 + I_{o}^2}
\end{equation}
The penalty term is :
\begin{equation}
	\label{deqn_ex20}
	L_{current} = E[max(0,I_{total} - 1.2*I_{peak})]
\end{equation}
Here, the $\max(0, \cdot)$ operation is the ReLU function which  ensures only excess currents are penalized. The ReLU (Rectified Linear Unit) function is a popular activation function in neural networks. It outputs zero for negative inputs and the input value itself for positive inputs, enabling efficient learning of complex patterns.
\item\textbf{ROCOF(Rate of change of frequency) loss}: To minimize the ROCOF
\begin{equation}
	\label{deqn_ex21}
	L_{ROCOF} = 1/N \sum_{i=1}^{N}|X_{i,R}|
\end{equation}
where, $X_{i,R}$ is the ROCOF input feature.
\item\textbf{ROCOV(Rate of change of voltage) loss}: To minimize the ROCOV
\begin{equation}
	\label{deqn_ex22}
	L_{ROCOV} = 1/N \sum_{i=1}^{N}|X_{i,V}|
\end{equation}
where, $X_{i,V}$ is the ROCOV input feature.
\end{itemize}
\begin{figure*}[htbp]
	\centering{\includegraphics[width=12.5cm,height=6cm]{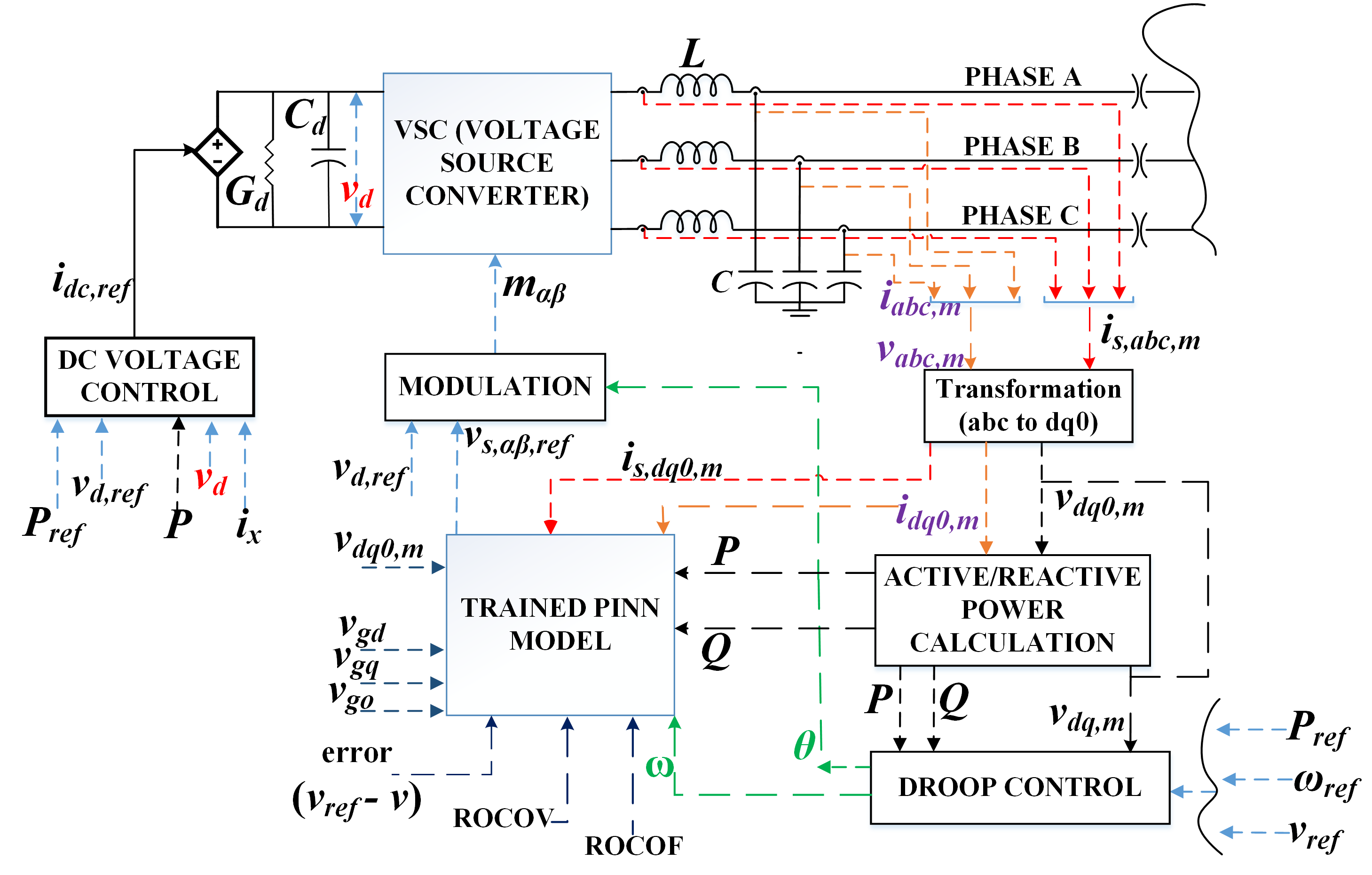}}
	\caption{Block diagram of proposed PINN-based GFC control}
	\label{fig.4}
\end{figure*}
So, the total loss is the combination of all the loss components and given by \textbf{$\mathcal{L}$} as 
\begin{equation}
    \label{deqn_ex23}
    \begin{aligned}
        \mathcal{L} = L_{MSE} &+ \lambda_{PDE} L_{Physics} + \lambda_{current} L_{current} \\
        &+ \lambda_{ROCOF} L_{ROCOF} + \lambda_{ROCOV} L_{ROCOV}
    \end{aligned}
\end{equation}
where $\lambda_{PDE}$, $\lambda_{current}$, $\lambda_{ROCOF}$ and $\lambda_{ROCOV}$ are hyperparameters controlling the weight of each term.
\subsubsection{Backpropagation and optimization}
The optimizer updates weights using the Adam optimizer, which does gradient calculation, moment estimation, Bias Correlation, and parameter update\cite{ref29}.
\subsubsection{PINN-based GFC control}
The proposed PINN-based GFC control strategy is shown in Fig. \ref{fig.4}. The trained PINN model replaces the AC voltage control, AC current limitation, and AC current control while retaining the classical frequency droop mechanism.
\section{RESULTS \& DISCUSSIONS}
\subsection{Simulation setup}
\begin{figure}[htbp]
	\centering{\includegraphics[width=10.5cm,height=6.0cm]{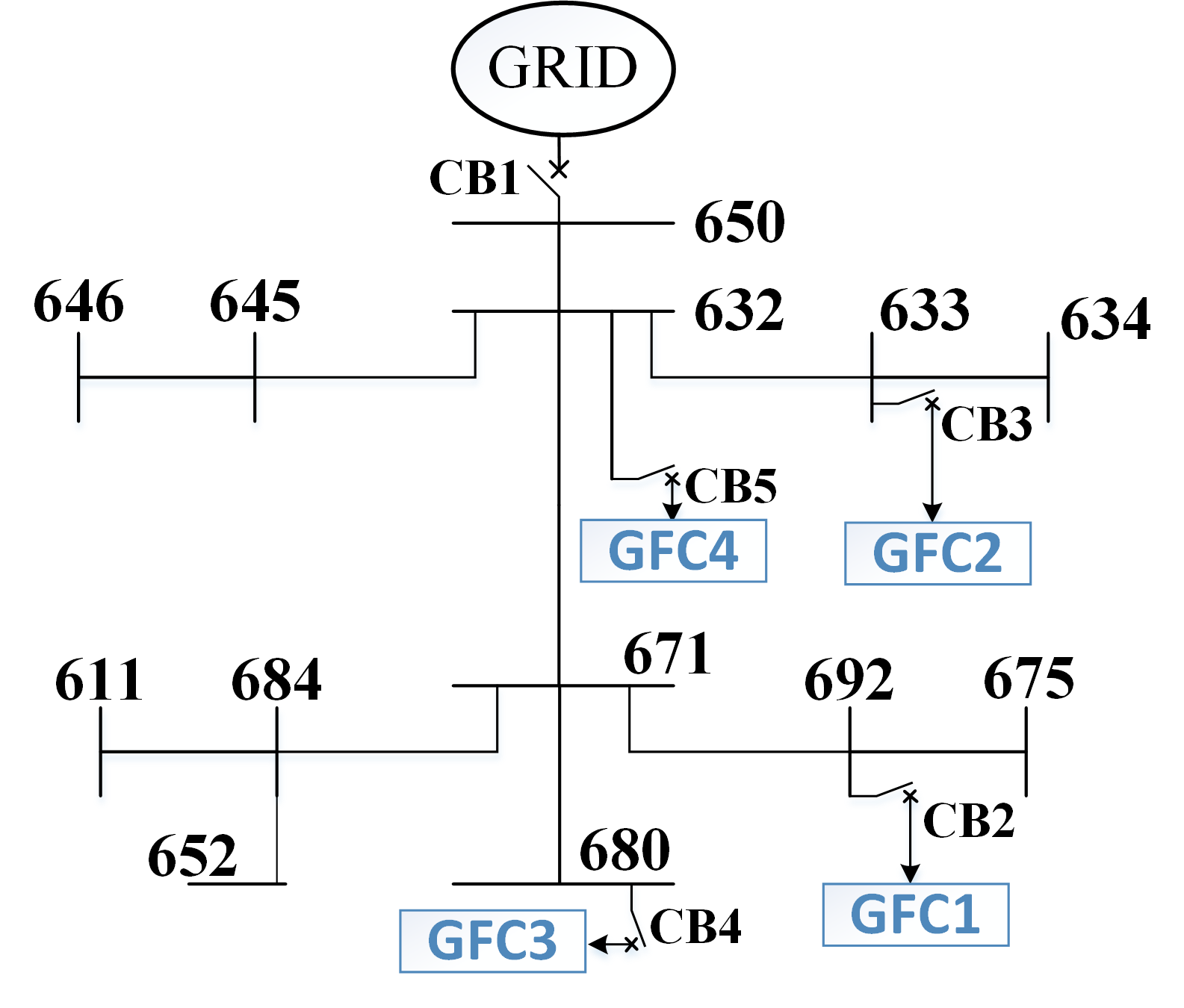}}
	\caption{Modified IEEE 13-Bus Test Feeder}
	\label{fig.5}
\end{figure}

\begin{figure}[htbp]
	\centering{\includegraphics[width=10.5cm,height=8.0cm]{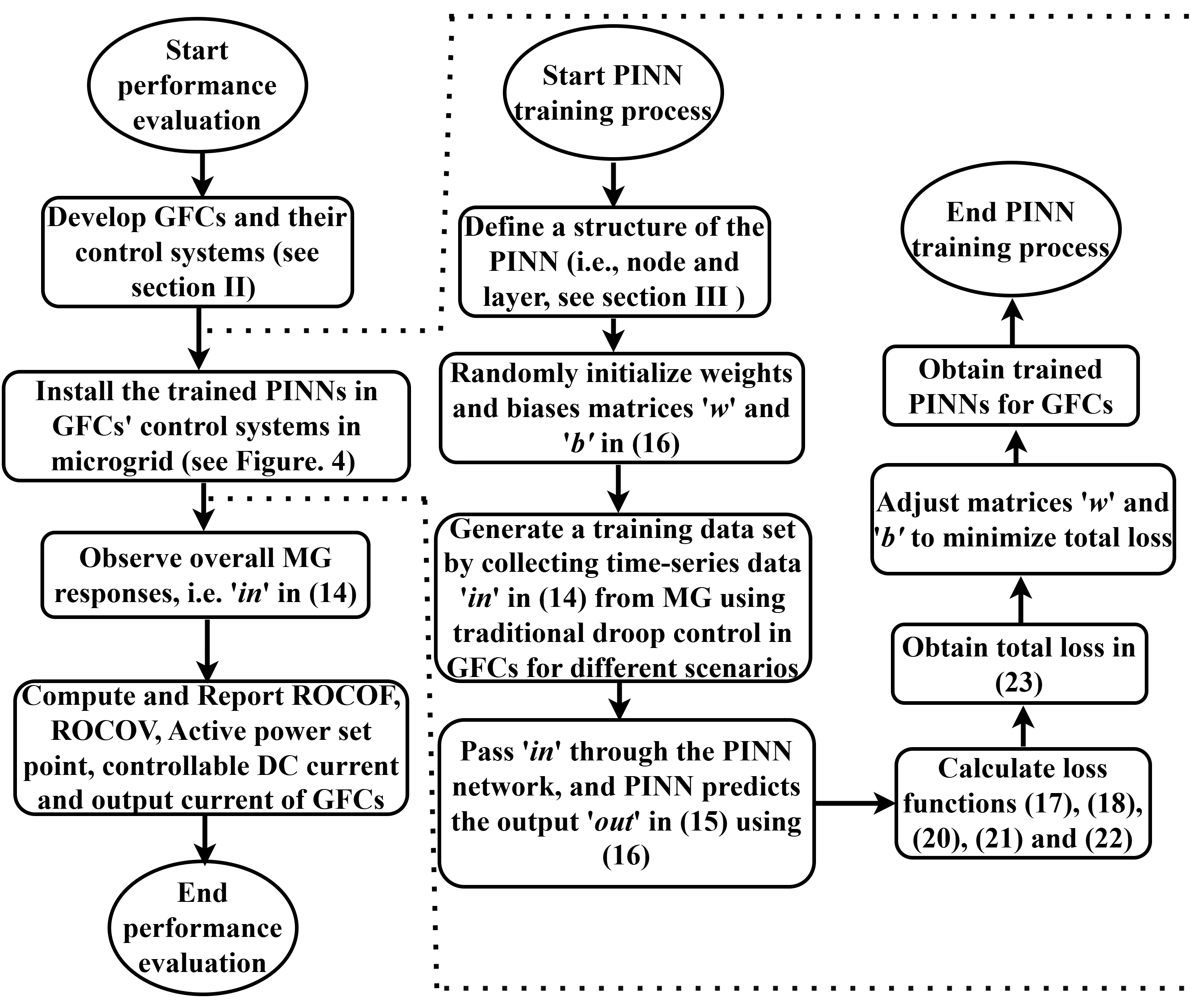}}
	\caption{Complete flowchart of the proposed PINN-based control strategy for GFCs}
	\label{fig.6}
\end{figure}

The 13-bus microgrid that is the subject of the study is depicted in Fig. \ref{fig.5}. It is a modified version of the standard IEEE 13-bus test feeder, with its parameters adopted from \cite{ref28}. Further, the system loads are changed to match up to the requirements of the test environment. Here, $CB{i}$, where i= 1,...,5, represents the circuit breakers and $GFC{j}$ where j= 1,..,4, represents the GFCs. The circuit breaker CB1 is used to allow the MG to be isolated from the main grid and to operate on-grid. Circuit breakers CB2, CB3, CB4, and CB5 are used to connect and island GFCs to the 13-bus MG. As indicated in section II, the base of each GFC is set at 50 Hz and 1.5 MVA. GFCs are capable of providing protection against both AC current limitations and DC source saturation. For each GFC, the local loads in the islanded condition are set at 0.375 MW per GFC. The analysis of the simulation results is divided into three parts for each case study:
 \begin{itemize}
    \item \textbf{First}, CB1 is closed, while CB2, CB3, CB4, and CB5 remain open. In this configuration, the GFCs supply power solely to their local loads, and the 13-bus microgrid receives power from the grid. This part analyzes the operation of GFCs in islanded mode, supplying power to their local loads. 
    \item  \textbf{Second}, with CB1 closed, CB2, CB3, CB4, and CB5 are closed at t = 0.8 seconds, allowing the GFCs to operate in on-grid mode. This section analyzes the transition of GFCs from islanded to on-grid operation. The grid parameters are used for control during on-grid operation of GFCs.
    \item  \textbf{Third}, with CB2, CB3, CB4, and CB5 closed, CB1 is opened at t = 5.5 seconds, causing the GFCs to supply power not only to their local loads but also to the 13-bus microgrid loads. This section examines the GFCs' capability to handle a sudden increase in power demand as they become the sole power source. During this scenario, factors such as DC source saturation and AC current limitation come into play due to the abrupt surge in power demand.
     
\end{itemize} 
MATLAB 2024b is used to implement the network in Fig. \ref{fig.5} in Simulink on a high-performance computer with a core i5-1245U, 1.60 GHz processor, and 16 GB of RAM. This environment is used for the PINN's training and simulation procedures as well as for the comparison strategies. The predict block in MATLAB Simulink is used to implement the trained PINN model. Wherever necessary, $\theta$ of (11) is used to change the coordinates for the correct input of the PINN. The PINN then uses (10) to anticipate the voltage reference, which further generates a modulation signal.\\
To minimize the loss function specified in (23), PINN is iteratively trained using 6000 iterations. The model is trained using simulation data that includes variations in both the local GFC load and the total load of the 13-bus microgrid (MG), along with the corresponding GFC responses under droop control as described in (11) and (12). For local loads, the variation occurs at random between 0\% and 90\% of the rated capacity of the GFC, and for the 13-bus MG load, it varies between 10\% and 100\% of the total rated capacity of all four GFCs. The data used for training the model is time series data with a fixed time step of 10 milliseconds, obtained from the system simulation shown in Fig. \ref{fig.5}, which runs for 9 seconds. The complete flowchart of the proposed PINN-based control strategy for GFCs is presented in Fig. \ref{fig.6}.
\subsection{Test results in modified IEEE 13-Bus Feeder}
\subsubsection{\underline{Case study 1}}
\begin{figure}[htbp]
	\centering{\includegraphics[width=12.5cm,height=4.0cm]{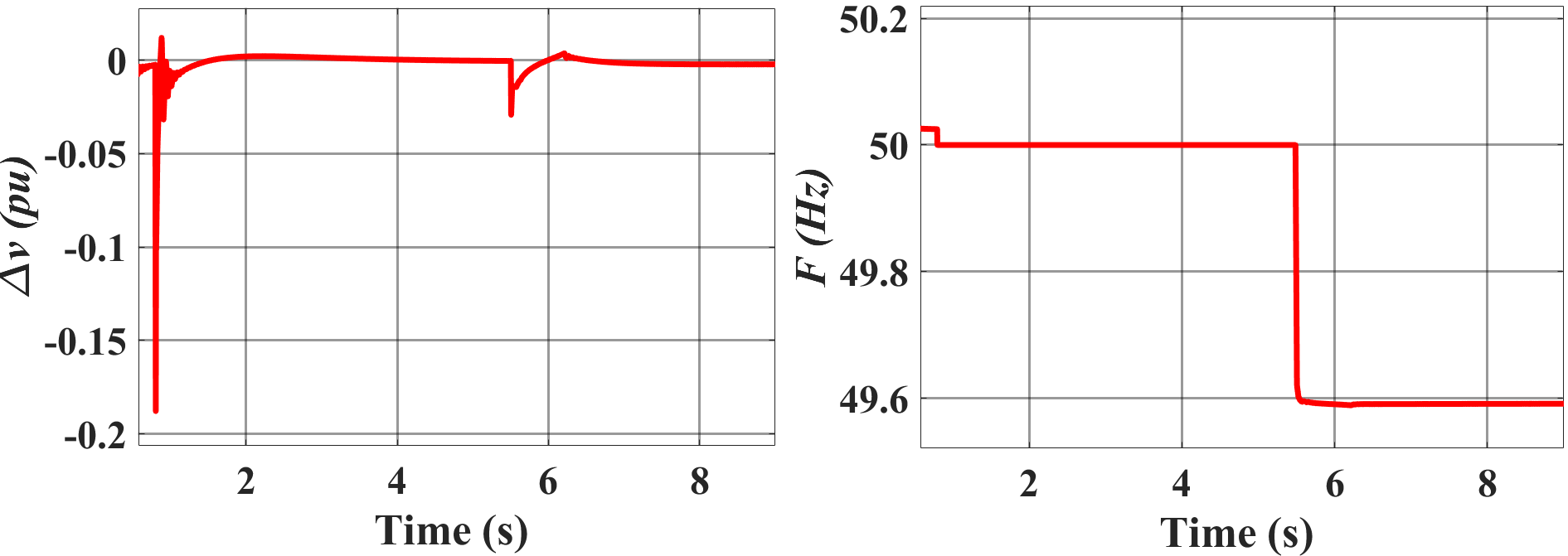}}
	\caption{Deviation of voltage and frequency of GFC1 for case study 1 at MG load of 5.94 MVA at 0.97 power factor}
	\label{fig.7}
\end{figure}

\begin{figure}[htbp]	\centering{\includegraphics[width=12.5cm,height=4.0cm]{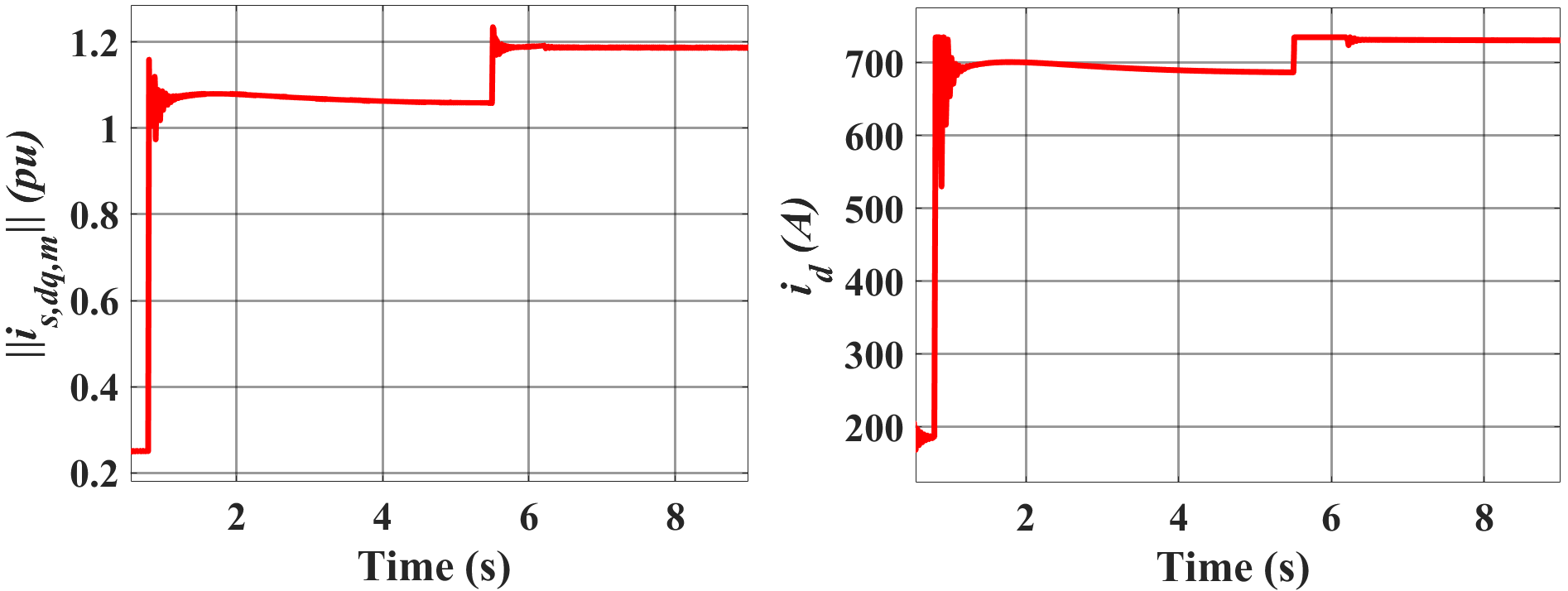}}
	\caption{Magnitude of output current and controllable DC current of GFC1 for case study 1 at MG load of 5.94 MVA at 0.97 power factor}
	\label{fig.8}
\end{figure}

\begin{figure}[htbp]
	\centering{\includegraphics[width=12.5cm,height=4.0cm]{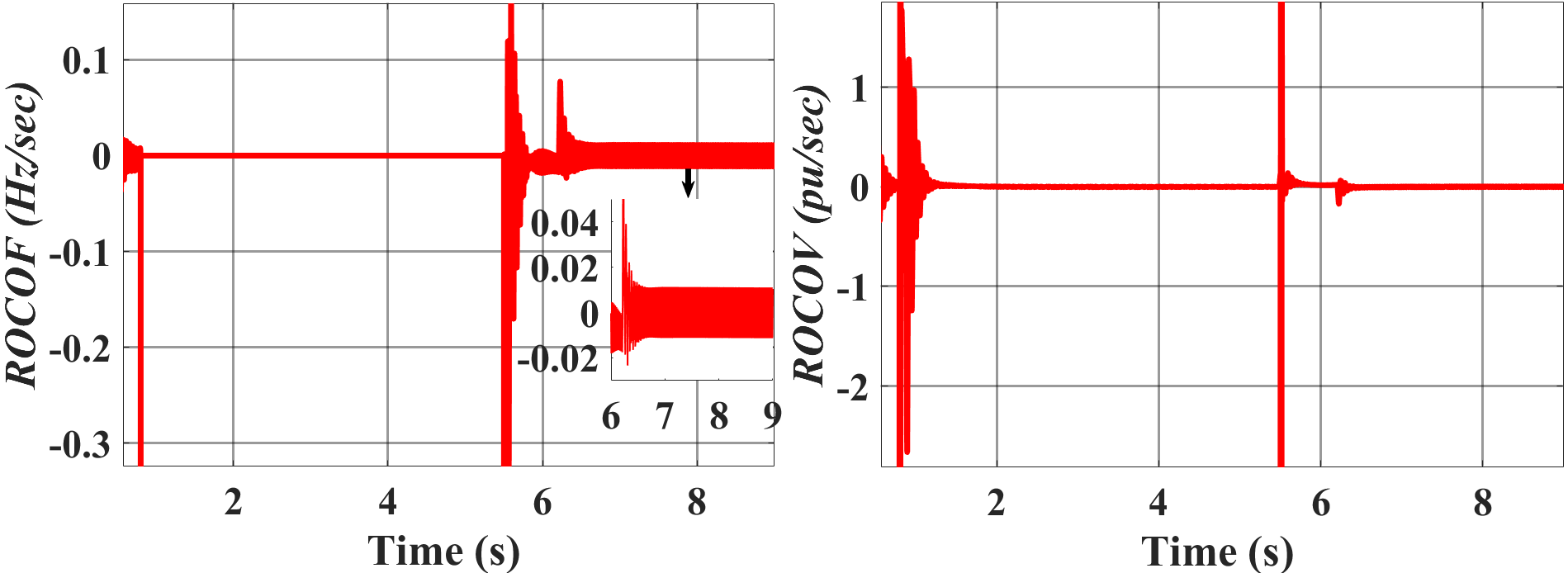}}
	\caption{ROCOF and ROCOV of GFC1 for case study 1 at MG load of 5.94 MVA at 0.97 power factor}
	\label{fig.9}
\end{figure}

The performance of the traditional droop control in the system (i.e., Fig. \ref{fig.5}) is evaluated in this case study. As illustrated in Fig. \ref{fig.7}, the GFC maintains stability with a microgrid (MG) load of 5.94 MVA at 0.97 power factor and a local load of 0.375 MW per GFC. Under these conditions, the GFC operates stably in both grid-connected and islanded modes as shown in Fig. \ref{fig.7}. The magnitude of the output current and the controllable DC current of GFC1, as shown in Fig. \ref{fig.8}, indicate that the GFC operates within its AC current limits. The ROCOF and ROCOV plots for this case are shown in Fig. \ref{fig.9}.
\begin{figure}[htbp]
	\centering{\includegraphics[width=12.5cm,height=4.0cm]{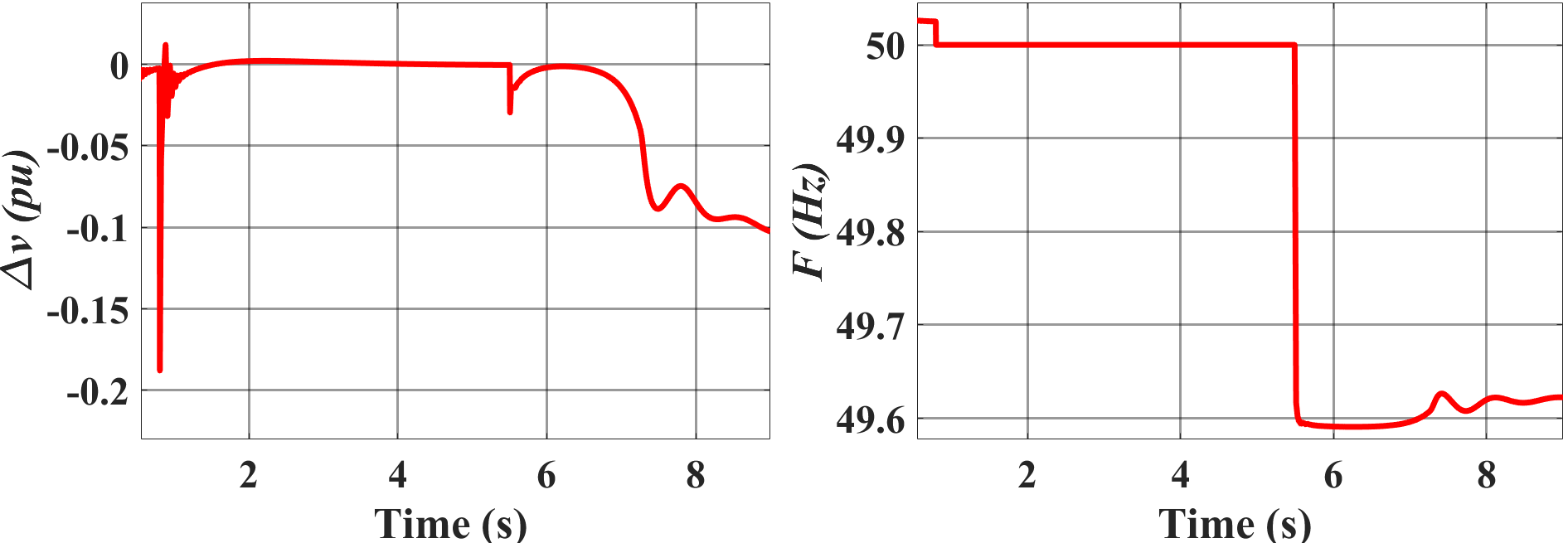}}
	\caption{Deviation of voltage and frequency of GFC1 for case study 1 at MG load of 5.942 MVA at 0.97 power factor}
	\label{fig.10}
\end{figure}
\begin{figure}[htbp]
	\centering{\includegraphics[width=12.5cm,height=4.0cm]{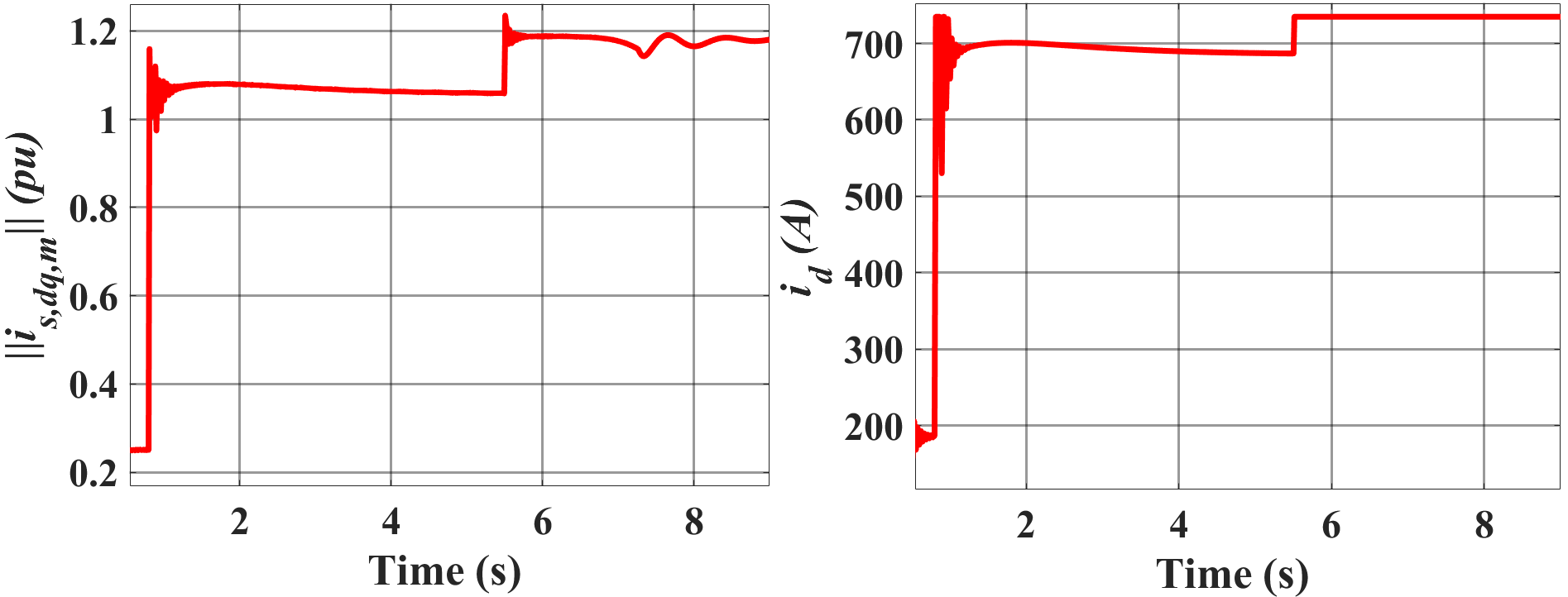}}
	\caption{Magnitude of output current and controllable DC current of GFC1 for case study 1 at MG load of 5.942 MVA at 0.97 power factor}
	\label{fig.11}
\end{figure}
\begin{figure}[htbp]
	\centering{\includegraphics[width=12.5cm,height=4cm]{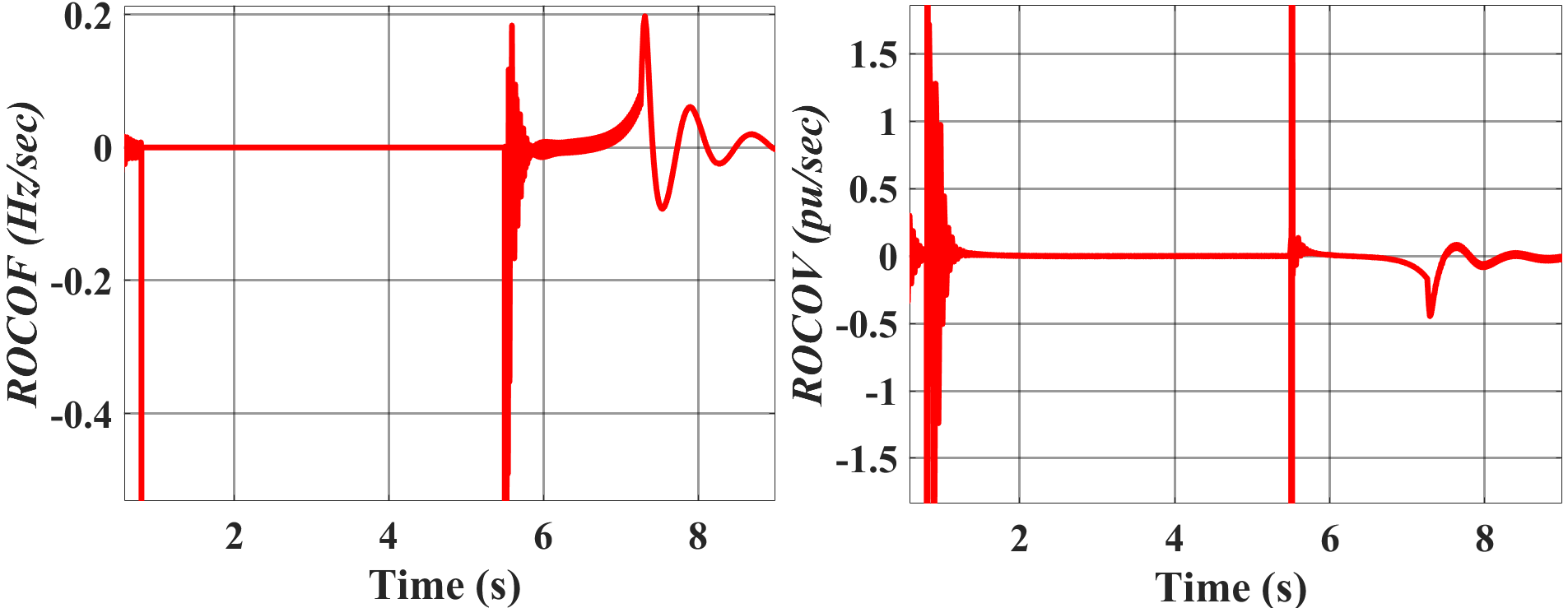}}
	\caption{ROCOF and ROCOV of GFC1 for case study 1 at MG load of 5.942 MVA at 0.97 power factor}
	\label{fig.12}
\end{figure} 
 However, when the load is slightly increased to an MG load of 5.942 MVA at 0.97 power factor (with the local load remaining at 0.375 MW per GFC), the GFC becomes unstable in island mode after t=5.5 seconds, as shown in Fig. \ref{fig.10}, due to the fixed set point of active power. The performance of the GFC begins to suffer under this load, as shown in Fig. \ref{fig.10}. The GFC1 voltage reduces to below 5\% of the steady state voltage for this load. In our study, the result of the GFC with the worst performance is displayed. Fig.\ref{fig.10} illustrates how the voltage drops for GFC1. Fig. \ref{fig.11} shows that under islanded conditions, the current flowing out of the controlled DC current source, $i_{d}$, reaches its maximum limit, $i^{d}_{max}$, at this load, even though the AC current limit, $i_{ac,max}$, is not exceeded. Fig. \ref{fig.12} illustrates the ROCOV and ROCOF for GFC1 in this scenario, showing a noticeable deterioration under islanded conditions compared to Fig.\ref{fig.9} (which shows ROCOF and ROCOV of GFC1 in stable condition).
\subsubsection{\underline{Case study 2}}
\begin{figure}[htbp]
	\centering{\includegraphics[width=12.5cm,height=4.0cm]{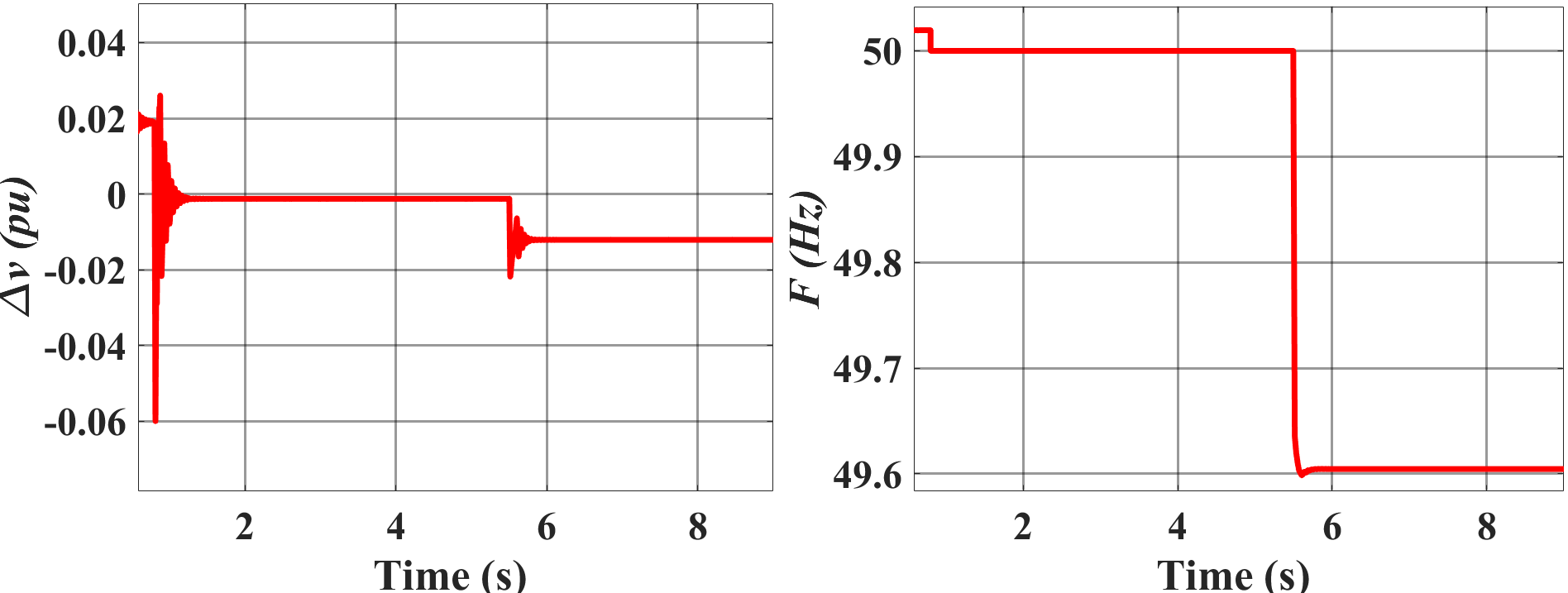}}
	\caption{Deviation of voltage and frequency of GFC1 for case study 2}
	\label{fig.13}
\end{figure}

To address the stability issue observed in Case Study 1, a PINN-based control strategy is implemented in addition to frequency droop control. The loading conditions remain unchanged, where the droop control failed to stabilize the microgrid during islanding. As shown in Fig. \ref{fig.13}, GFC1 successfully transitions from the off-grid to the on-grid mode at t=0.8 seconds and returns to the islanded mode at t=5.5 seconds. In particular, PINN-based control effectively stabilizes the system even with an MG load of 5.942 MVA at 0.97 power factor and a local load of 0.375 MW per GFC in the islanded mode. The voltage deviation also remains within acceptable 5\% of the steady state voltage for this load.
\begin{figure}[htbp]
	\centering{\includegraphics[width=12.5cm,height=4.0cm]{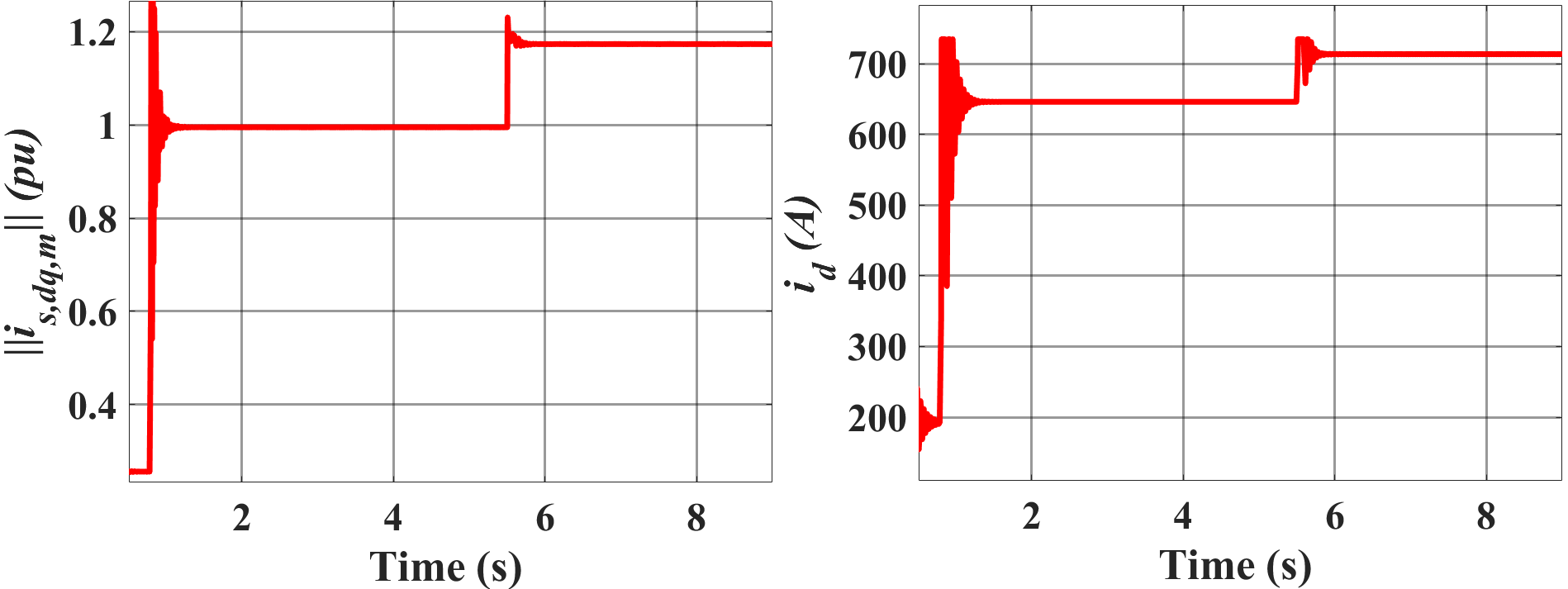}}
	\caption{Magnitude of output current and controllable DC current of GFC1 for case study 2}
	\label{fig.14}
\end{figure}
\begin{figure}[htbp]
	\centering{\includegraphics[width=12.5cm,height=4.0cm]{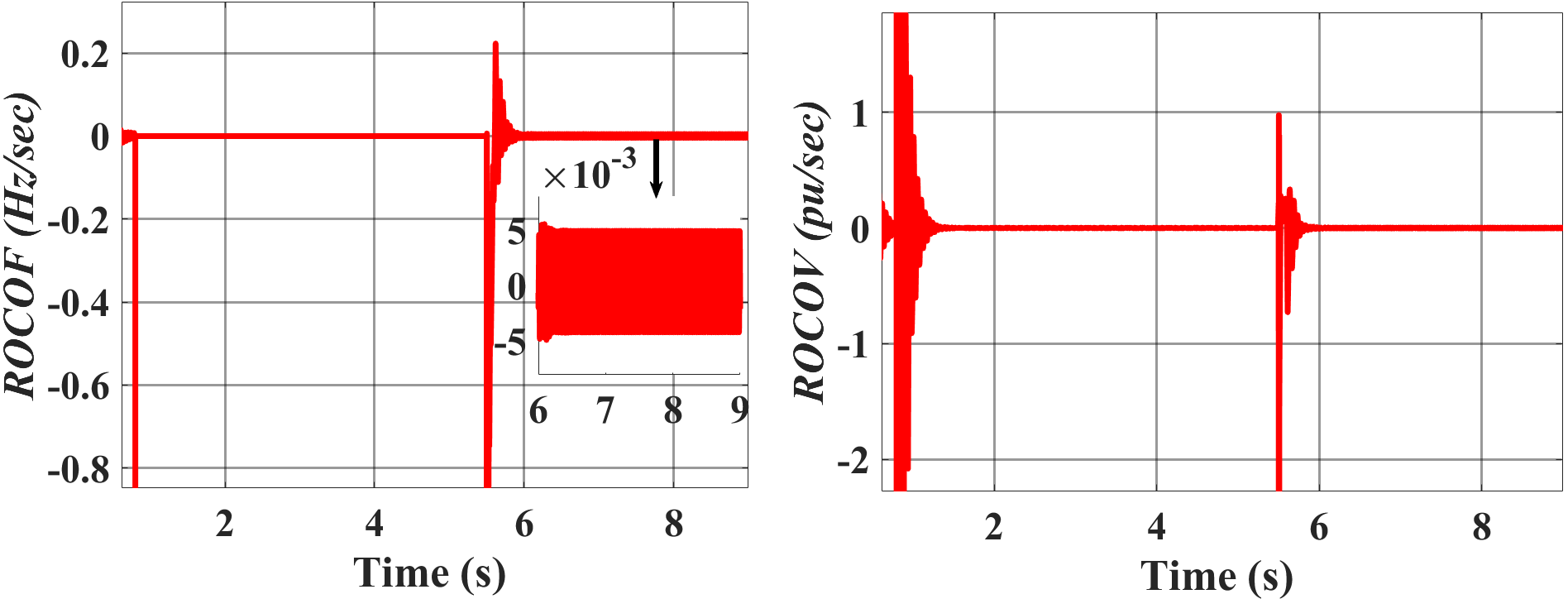}}
	\caption{ROCOF and ROCOV of GFC1 for case study 2}
	\label{fig.15}
\end{figure}
As seen in Fig. \ref{fig.14}, in this case, $i_d$ has not reached the maximum limit $i^{d}_{max}$ in the islanded situation. Peak voltage deviation observed during transient reduced to 24.14\% (i.e., from -0.029 to -0.022 as shown in Fig. \ref{fig.7} and Fig. \ref{fig.13}) in this case study than in case study 1, and the MG is also successfully stabilized. Fig. \ref{fig.15} presents the ROCOF and ROCOV for this case, demonstrating improved performance over Case Study 1 (Fig. \ref{fig.9}), with ROCOF improved from 0.02 Hz/sec to 0.005 Hz/sec. 
\subsubsection{\underline{Case study 3}}
\begin{figure}[htbp]
	\centering{\includegraphics[width=12.5cm,height=4.0cm]{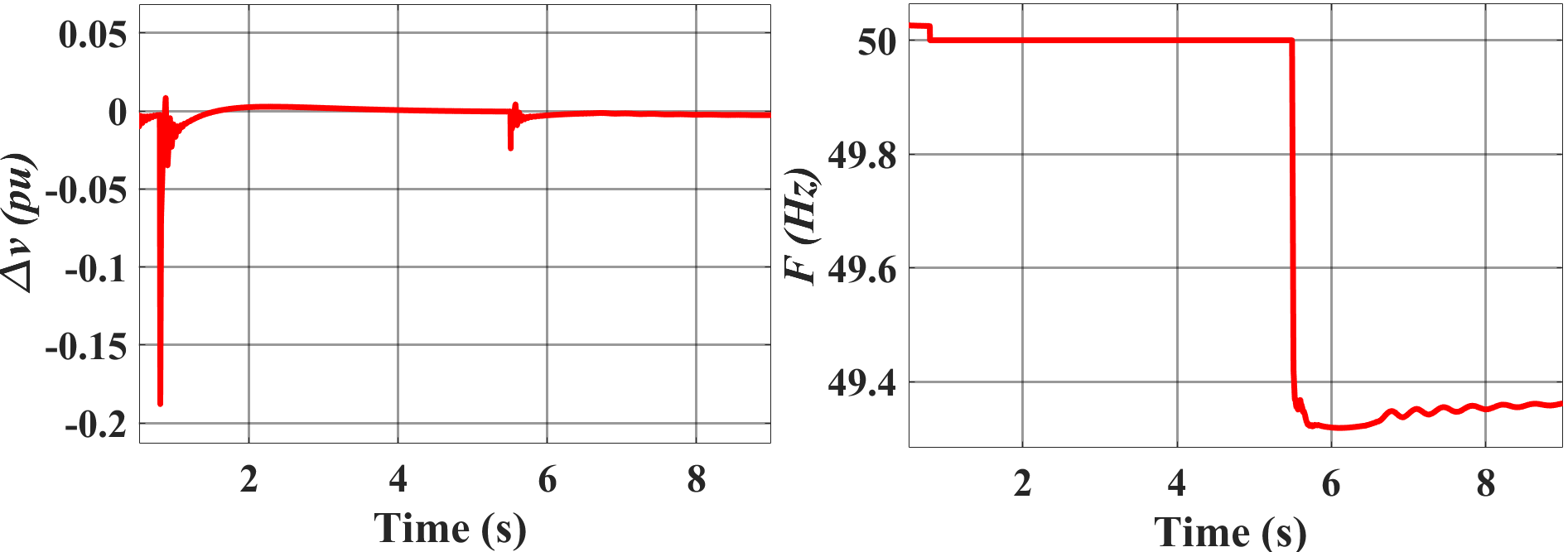}}
	\caption{Deviation of voltage and frequency of GFC1 using current limitation strategy of \cite{ref11}}
	\label{fig.16}
\end{figure}
\begin{figure}[htbp]
	\centering{\includegraphics[width=12.5cm,height=4.0cm]{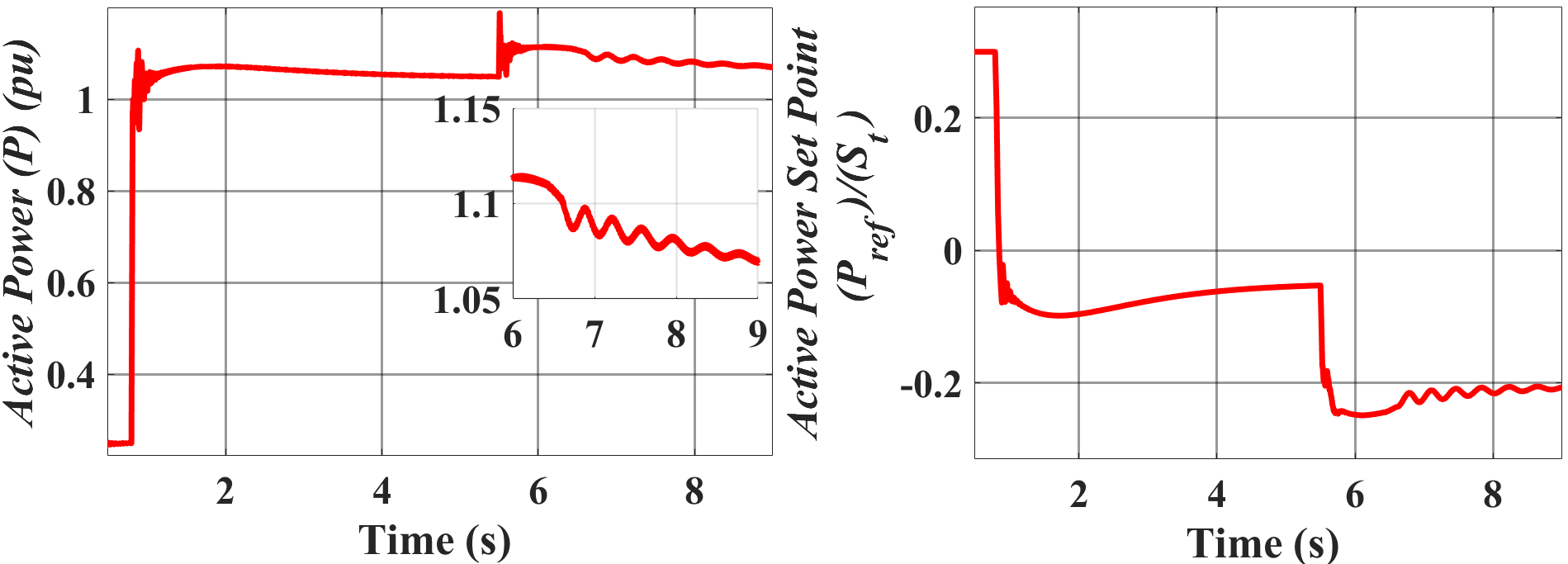}}
	\caption{Active power and Active power set point of GFC1 using current limitation strategy of \cite{ref11}}
	\label{fig.17}
\end{figure}
\begin{figure}[htbp]
	\centering{\includegraphics[width=12.5cm,height=4.0cm]{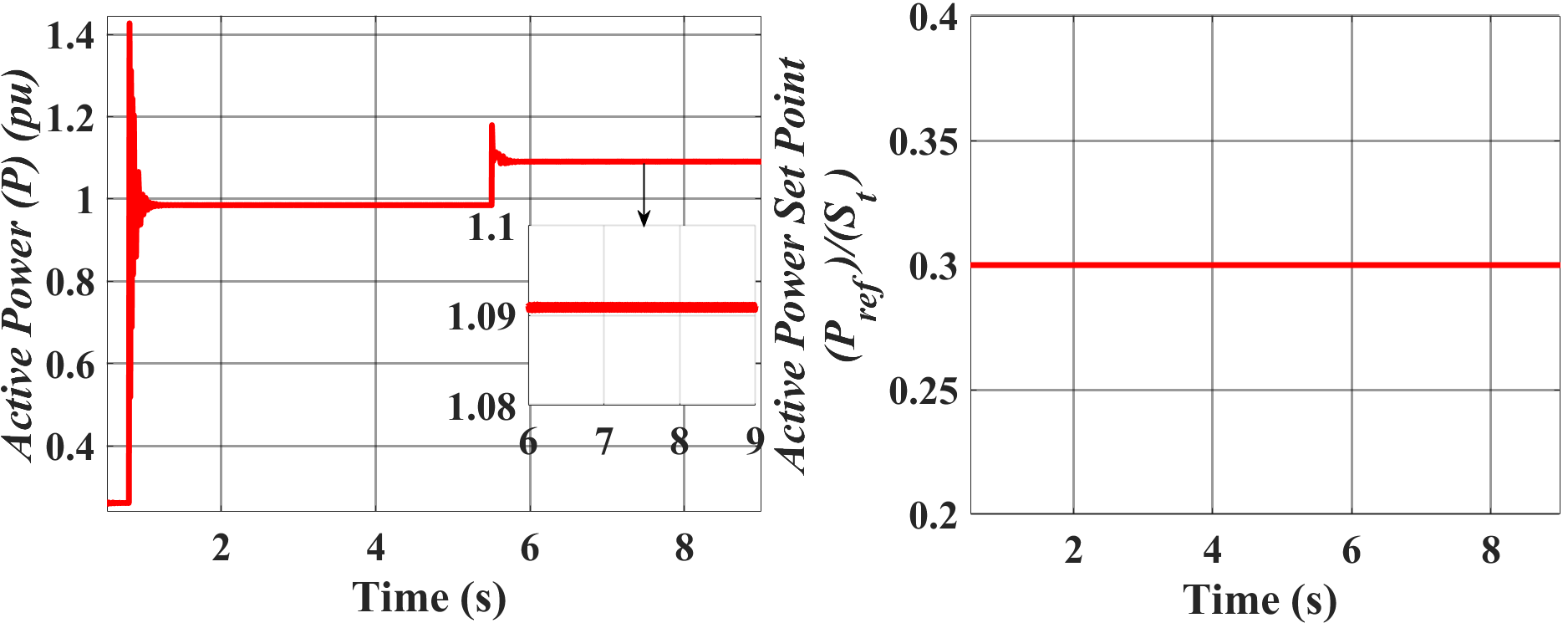}}
	\caption{Active power and Active power set point of GFC1 using PINN based control strategy}
	\label{fig.18}
\end{figure}
This case study compares the current limitation strategy proposed in \cite{ref11} with the PINN-based control strategy introduced in this paper. Both approaches successfully stabilized the system under an MG load of 5.942 MVA at 0.97 power factor and a local load of 0.375 MW per GFC in island mode, as shown in Fig. \ref{fig.13} and Fig. \ref{fig.16}.
However, a key difference arises in preserving the post-disturbance operating point of the GFC, as illustrated in Fig. \ref{fig.17} and Fig. \ref{fig.18}. The PINN-based control effectively stabilizes the system without altering the post-disturbance operating point as shown in Fig. \ref{fig.18}.
 In contrast, the current limitation strategy proposed in \cite{ref11} modifies the post-disturbance operating point since it relies on predefined threshold values to adjust the active power setpoint as shown in Fig. \ref{fig.17}. The reliance on predefined threshold values presents a drawback: even during normal operation (at a load that consumes an AC current greater than 0.9 p.u. of the rated AC current), this strategy adjusts the active power setpoint. 
 The PINN-based control improves the frequency of the system by up to 0.245 Hz (i.e., 49.605 Hz - 49.36 Hz), as shown in Fig. \ref{fig.13} and Fig. \ref{fig.16}. Also, the PINN-based control improves the output active power of GFC1 by up to 0.03 p.u. (i.e., 1.09 p.u. - 1.06 p.u.), as shown in Fig. \ref{fig.17} and Fig. \ref{fig.18}. 
\subsubsection{\underline{Case study 4}}
\begin{figure}[htbp]	 
    \centering{\includegraphics[width=12.5cm,height=4.0cm]{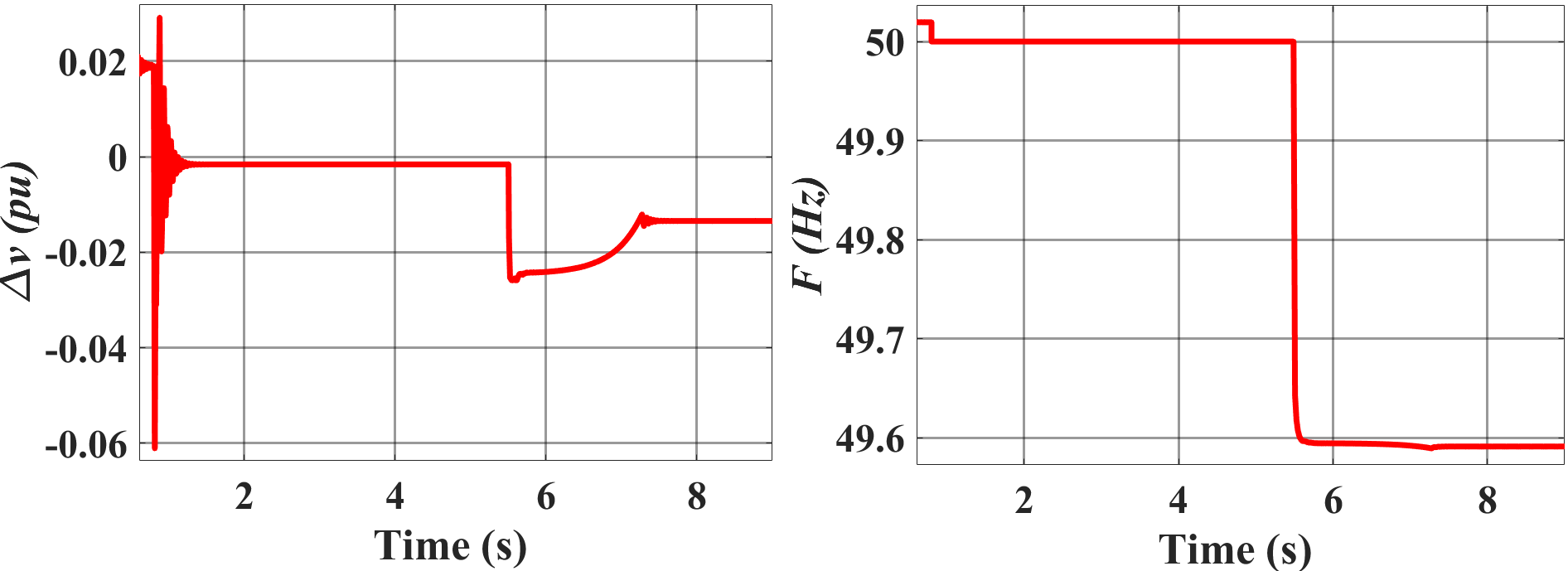}}
    \caption{Deviation of voltage and frequency of GFC1 for case study 3}
	\label{fig.19}
\end{figure} 
\begin{figure}[htbp]
	\centering{\includegraphics[width=12.5cm,height=4.0cm]{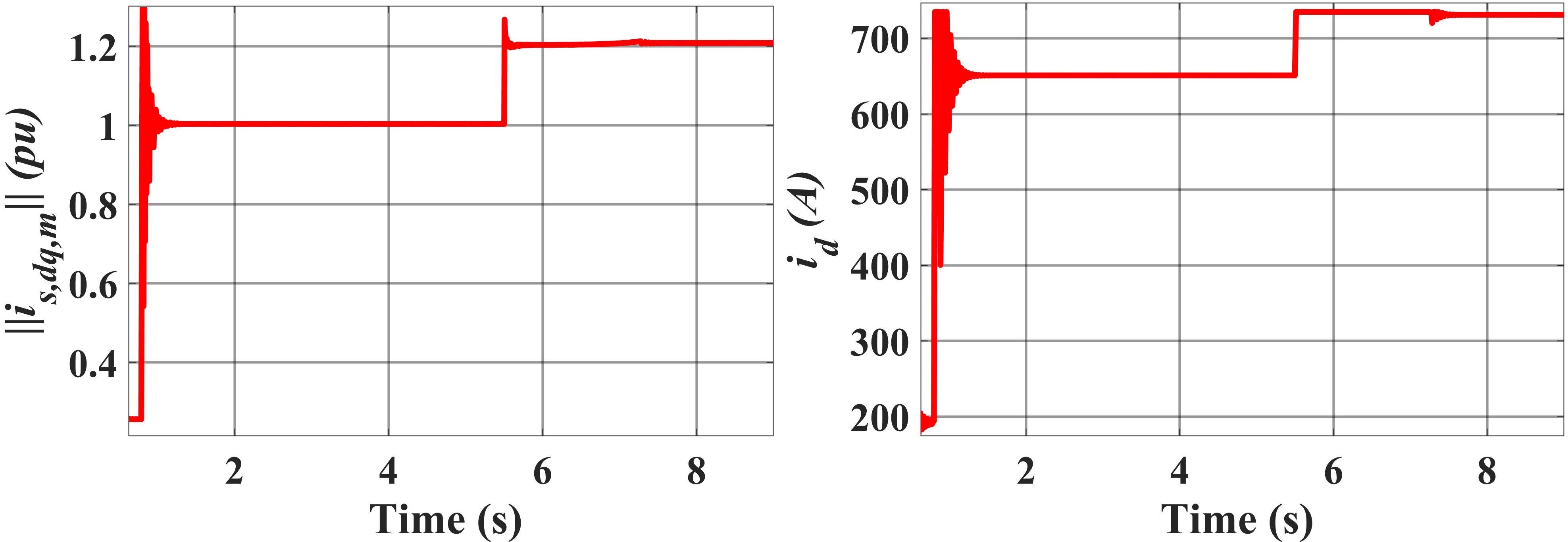}}
	\caption{Magnitude of output current and controllable DC current of GFC1 for case study 3}
	\label{fig.20}
\end{figure}
This case study evaluates the effectiveness of the PINN-based control strategy for GFCs in maintaining stability under maximum sudden load changes. Through multiple simulation scenarios, it was observed that the GFCs remained stable with PINN-based control up to an MG load of 6.197 MVA at 0.97 power factor and a local load of 0.375 MW per GFC, as shown in Fig. \ref{fig.19}. At this load level, the GFC1 reaches both its AC current limit and controlled DC current limit, as shown in Fig. \ref{fig.20}. Beyond this point, further load increases result in system instability, even with the PINN-based control in place. PINN can withstand a load increase of 6.197 MVA at 0.97 power factor from 5.94 MVA at 0.97 power factor compared to traditional droop control without compromising system stability, as shown in Fig. \ref{fig.19}.
\par The results demonstrate that GFCs operating under traditional droop control, even when equipped with protection mechanisms for AC current limitations and DC source saturation, fail to ensure system stability under sudden overloading conditions. In contrast, the proposed PINN-based control strategy successfully stabilizes the system where traditional droop control proves inadequate, all while maintaining compliance with AC current and DC source protection constraints. Furthermore, the study identifies the maximum load threshold up to which the PINN-based controller can effectively sustain system stability during abrupt load changes, without compromising protection against AC current and DC source saturation.
\section{Conclusion}
This work presents a physics-informed neural network (PINN)-based control strategy for grid-forming converters (GFCs) aimed at enhancing microgrid (MG) stability. The proposed approach is developed to address the limitations of traditional droop control methods, particularly under challenging operating conditions such as DC source saturation and AC current limitation.
A key research gap addressed in this study is the instability introduced by conventional droop control when current limitations are enforced in GFCs. In such scenarios, maintaining a fixed active power setpoint across varying loads can lead to control instability and degraded performance, especially under overload conditions.
The proposed PINN-based controller offers several advantages over traditional methods. It significantly improves voltage and frequency stability, demonstrating enhanced performance during both steady-state and transient conditions while maintaining a fixed active power setpoint across varying loads. Notably, it enables robust operation during MG islanding and ensures smooth transitions with minimal disturbances during grid synchronization. By reducing transient responses and supporting seamless resynchronization, the PINN contributes to more resilient and adaptive MG operation.
\bibliography{ref}

\begin{thebibliography}{10}
\expandafter\ifx\csname url\endcsname\relax
  \def\url#1{\texttt{#1}}\fi
\expandafter\ifx\csname urlprefix\endcsname\relax\def\urlprefix{URL }\fi
\expandafter\ifx\csname href\endcsname\relax
  \def\href#1#2{#2} \def\path#1{#1}\fi

\bibitem{ref1}
D.~Pattabiraman, R.~H. Lasseter, T.~M. Jahns, Comparison of grid following and grid forming control for a high inverter penetration power system, in: 2018 IEEE Power \& Energy Society General Meeting (PESGM), IEEE, 2018, pp. 1--5.

\bibitem{ref2}
M.~C. Chandorkar, D.~M. Divan, R.~Adapa, Control of parallel connected inverters in standalone ac supply systems, IEEE transactions on industry applications 29~(1) (2002) 136--143.

\bibitem{ref3}
N.~Pogaku, M.~Prodanovic, T.~C. Green, Modeling, analysis and testing of autonomous operation of an inverter-based microgrid, IEEE Transactions on power electronics 22~(2) (2007) 613--625.

\bibitem{ref4}
Q.-C. Zhong, G.~Weiss, Synchronverters: Inverters that mimic synchronous generators, IEEE transactions on industrial electronics 58~(4) (2010) 1259--1267.

\bibitem{ref5}
Z.~Kustanovich, S.~Shivratri, H.~Yin, F.~Reissner, G.~Weiss, Synchronverters with fast current loops, IEEE Transactions on Industrial Electronics 70~(11) (2022) 11357--11367.

\bibitem{ref6}
S.~D'Arco, J.~A. Suul, Equivalence of virtual synchronous machines and frequency-droops for converter-based microgrids, IEEE Transactions on Smart Grid 5~(1) (2013) 394--395.

\bibitem{ref7}
S.~V. Dhople, B.~B. Johnson, A.~O. Hamadeh, Virtual oscillator control for voltage source inverters, in: 2013 51st annual allerton conference on communication, control, and computing (Allerton), IEEE, 2013, pp. 1359--1363.

\bibitem{ref8}
B.~B. Johnson, M.~Sinha, N.~G. Ainsworth, F.~D{\"o}rfler, S.~V. Dhople, Synthesizing virtual oscillators to control islanded inverters, IEEE Transactions on Power Electronics 31~(8) (2015) 6002--6015.

\bibitem{ref9}
D.~Gro{\ss}, M.~Colombino, J.-S. Brouillon, F.~D{\"o}rfler, The effect of transmission-line dynamics on grid-forming dispatchable virtual oscillator control, IEEE Transactions on Control of Network Systems 6~(3) (2019) 1148--1160.

\bibitem{ref10}
C.~Arghir, T.~Jouini, F.~D{\"o}rfler, Grid-forming control for power converters based on matching of synchronous machines, Automatica 95 (2018) 273--282.

\bibitem{ref11}
A.~Tayyebi, D.~Gro{\ss}, A.~Anta, F.~Kupzog, F.~D{\"o}rfler, Frequency stability of synchronous machines and grid-forming power converters, IEEE Journal of Emerging and Selected Topics in Power Electronics 8~(2) (2020) 1004--1018.

\bibitem{ref12}
K.~Strunz, K.~Almunem, C.~Wulkow, M.~Kuschke, M.~Valescudero, X.~Guillaud, Enabling 100\% renewable power systems through power electronic grid-forming converter and control: System integration for security, stability, and application to europe, Proceedings of the IEEE 111~(7) (2022) 891--915.

\bibitem{ref13}
S.~Samanta, N.~R. Chaudhuri, C.~M. Lagoa, Fast frequency support from grid-forming converters under dc-and ac-side current limits, IEEE Transactions on Power Systems 38~(4) (2022) 3528--3542.

\bibitem{ref14}
D.~B. Rathnayake, B.~Bahrani, Multivariable control design for grid-forming inverters with decoupled active and reactive power loops, IEEE Transactions on Power Electronics 38~(2) (2022) 1635--1649.

\bibitem{ref15}
A.~Arjomandi-Nezhad, Y.~Guo, B.~C. Pal, D.~Varagnolo, A model predictive approach for enhancing transient stability of grid-forming converters, IEEE Transactions on Power Systems (2024).

\bibitem{ref16}
G.~W. Chang, K.~T. Nguyen, A new adaptive inertia-based virtual synchronous generator with even inverter output power sharing in islanded microgrid, IEEE Transactions on Industrial Electronics 71~(9) (2023) 10693--10703.

\bibitem{ref17}
M.~Mousavizade, F.~Bai, R.~Garmabdari, M.~Sanjari, F.~Taghizadeh, A.~Mahmoudian, J.~Lu, Adaptive control of v2gs in islanded microgrids incorporating ev owner expectations, Applied Energy 341 (2023) 121118.

\bibitem{ref18}
P.~Khemmook, K.~Prompinit, T.~Surinkaew, Control of a microgrid using robust data-driven-based controllers of distributed electric vehicles, Electric Power Systems Research 213 (2022) 108681.

\bibitem{ref19}
H.~Abubakr, A.~Lashab, J.~C. Vasquez, T.~H. Mohamed, J.~M. Guerrero, Novel v2g regulation scheme using dual-pss for pv islanded microgrid, Applied Energy 340 (2023) 121012.

\bibitem{ref20}
Y.~Shan, J.~Hu, K.~W. Chan, S.~Islam, A unified model predictive voltage and current control for microgrids with distributed fuzzy cooperative secondary control, IEEE Transactions on Industrial Informatics 17~(12) (2021) 8024--8034.

\bibitem{ref21}
H.~R. Baghaee, M.~Mirsalim, G.~B. Gharehpetian, Performance improvement of multi-der microgrid for small-and large-signal disturbances and nonlinear loads: Novel complementary control loop and fuzzy controller in a hierarchical droop-based control scheme, IEEE Systems Journal 12~(1) (2016) 444--451.

\bibitem{ref22}
I.~Ngamroo, T.~Surinkaew, Control of distributed converter-based resources in a zero-inertia microgrid using robust deep learning neural network, IEEE Transactions on Smart Grid 15~(1) (2023) 49--66.

\bibitem{ref23}
J.~Kaushal, P.~Basak, Power quality control based on voltage sag/swell, unbalancing, frequency, thd and power factor using artificial neural network in pv integrated ac microgrid, Sustainable Energy, Grids and Networks 23 (2020) 100365.

\bibitem{ref25}
I.~Ngamroo, T.~Surinkaew, Resiliency-guided grid-forming converter control of distributed solar-powered electric vehicles, IEEE Transactions on Intelligent Transportation Systems (2024).

\bibitem{ref24}
B.~Zhang, W.~Hu, X.~Xu, T.~Li, Z.~Zhang, Z.~Chen, Physical-model-free intelligent energy management for a grid-connected hybrid wind-microturbine-pv-ev energy system via deep reinforcement learning approach, Renewable Energy 200 (2022) 433--448.

\bibitem{ref26a}
H.~Xin, L.~Huang, L.~Zhang, Z.~Wang, J.~Hu, Synchronous instability mechanism of pf droop-controlled voltage source converter caused by current saturation, IEEE Transactions on Power Systems 31~(6) (2016) 5206--5207.

\bibitem{ref27}
M.~G. Taul, X.~Wang, P.~Davari, F.~Blaabjerg, Current limiting control with enhanced dynamics of grid-forming converters during fault conditions, IEEE Journal of Emerging and Selected Topics in Power Electronics 8~(2) (2019) 1062--1073.

\bibitem{ref29}
B.~Imtiaz, I.~Zafar, C.~Yuanhui, Modelling of an optimized microgrid model by integrating dg distributed generation sources to ieee 13 bus system, European Journal of Electrical Engineering and Computer Science 5~(2) (2021) 18--25.

\bibitem{ref28}
B.~Huang, J.~Wang, Applications of physics-informed neural networks in power systems-a review, IEEE Transactions on Power Systems 38~(1) (2022) 572--588.

\end{thebibliography}

\end{document}